\documentclass[acmsmall,screen,nonacm]{acmart}

\AtBeginDocument{%
  }

\authorsaddresses{}

\usepackage{amsmath,amsfonts,amsthm}
\usepackage{algorithmic}
\usepackage{graphicx}
\usepackage{lipsum}
\usepackage{wrapfig}
\usepackage{float}
\usepackage{textcomp}
\usepackage{colortbl}
\usepackage{xcolor}
\usepackage{xspace}
\usepackage{array}
\usepackage{soul}
\usepackage{url}
\usepackage{framed}
\usepackage{tcolorbox}
\tcbuselibrary{breakable}
\usepackage{hyperref}
\usepackage{listings}
\usepackage{tabularx}
\usepackage{makecell}
\usepackage{booktabs}
\usepackage{calc}
\usepackage{fancyvrb}
\usepackage{adjustbox}
\usepackage{multirow}
\usepackage{enumitem}
\usepackage{subcaption}
\usepackage{titlesec}
\definecolor{codegreen}{rgb}{0,0.6,0}
\definecolor{codegray}{rgb}{0.5,0.5,0.5}
\definecolor{codepurple}{rgb}{0.58,0,0.82}
\definecolor{backcolour}{rgb}{0.95,0.95,0.92}
\definecolor{brickred}{rgb}{0.8, 0.25, 0.33}
\definecolor{nightBlue}{RGB}{45,78,159}

\lstdefinestyle{mystyle}{
  morecomment=[f][\color{codegreen}]{+\ },
  morecomment=[f][\color{brickred}]{-\ },
  backgroundcolor=\color{backcolour},   
  commentstyle=\color{codegreen},
  keywordstyle=\color{magenta},
  numberstyle=\tiny\color{codegray},
  stringstyle=\color{codepurple},
  basicstyle=\linespread{1}\ttfamily\footnotesize,
  breakatwhitespace=false,         
  breaklines=true,                 
  captionpos=b,                    
  keepspaces=true,                 
  numbers=left,                    
  numbersep=3pt,                  
  showspaces=false,                
  showstringspaces=false,
  showtabs=false,                  
  tabsize=2
}
\setlist[itemize]{leftmargin=10pt}
\newcounter{findingcounter}
\newtcolorbox{finding}{
  breakable,
  colback=gray!30!white, 
  boxrule=.5pt, 
  left=.5mm, 
  right=.5mm, 
  top=.75mm, 
  bottom=.75mm, 
  boxsep=1mm, 
  arc=0mm, 
  outer arc=0mm, 
  before upper={
  \refstepcounter{findingcounter}
  \textbf{Finding \thefindingcounter: }
  }
}

\newcommand{\patternpara}[1]{\noindent\textbf{\textit{#1}\mbox{}.}}

\iftrue
\newcommand{\josh}[1]{{\color{red}(Josh: #1)}}

\newcommand{\yuntianyi}[1]{{\color{cyan}[Yuntianyi: #1]}}
\newcommand{\yuntianyihl}[1]{{\color{black}{#1}}}
\newcommand{\yuntianc}[1]{{\color{black}{#1}}}
\newcommand{\Yuqi}[1]{{\color{orange}[Yuqi: #1]}}

\else 
\newcommand{\josh}[1]{}
\newcommand{\yuntianyi}[1]{}
\newcommand{\Yuqi}[1]{}
\fi

\newcommand{\numsyn}{15\xspace}
\newcommand{\numsem}{27\xspace}
\newcommand{\numbug}{1,331\xspace}

\newcommand{\bfps}{bug-fix patterns\xspace}
\newcommand{\bfp}{bug-fix pattern\xspace}
\newcommand{\Bfps}{Bug-fix patterns\xspace}
\newcommand{\Bfp}{Bug-fix pattern\xspace}
\newcommand{\BFPs}{Bug-Fix Patterns\xspace}
\newcommand{\BFP}{Bug-Fix Pattern\xspace}

\newcommand{\bfas}{bug-fix actions\xspace}
\newcommand{\bfa}{bug-fix action\xspace}
\newcommand{\Bfas}{Bug-fix actions\xspace}
\newcommand{\Bfa}{Bug-fix action\xspace}
\newcommand{\BFAs}{Bug-Fix Actions\xspace}
\newcommand{\BFA}{Bug-Fix Action\xspace}

\newcommand{\lidar}{LiDAR\xspace}

\theoremstyle{definition}

\titlespacing\paragraph{0pt}{2pt plus 0pt minus 2pt}{4pt plus 2pt minus 1pt}

\expandafter\def\expandafter\normalsize\expandafter{%
    \normalsize
    \setlength\abovedisplayskip{3pt}
    \setlength\belowdisplayskip{2pt}
    \setlength\abovedisplayshortskip{5pt}
    \setlength\belowdisplayshortskip{5pt}
}

\makeatletter
\def\thm@space@setup{%
  \thm@preskip=0pt
  \thm@postskip=\thm@preskip 
}
\makeatother

\newcommand\blfootnote[1]{%
  \begingroup
  \renewcommand\thefootnote{}\footnote{#1}%
  \addtocounter{footnote}{-1}%
  \endgroup
}

\begin{document}
    \title{A Comprehensive Study of Bug-Fix Patterns in Autonomous Driving Systems}

    \author{Yuntianyi Chen}
    \orcid{0000-0002-3497-4167}
    \affiliation{%
      \institution{University of California, Irvine}
      \city{Irvine}
      \country{USA}
    }
    \email{yuntianc@uci.edu}
    
    \author{Yuqi Huai}
    \orcid{0000-0002-4792-8215}
    \affiliation{%
      \institution{University of California, Irvine}
      \city{Irvine}
      \country{USA}
    }
    \email{yhuai@uci.edu}

    \author{Yirui He}
    \orcid{0009-0009-6248-2053}
    \affiliation{%
      \institution{University of California, Irvine}
      \city{Irvine}
      \country{USA}
    }
    \email{yiruih@uci.edu}
    
    \author{Shilong Li}
    \orcid{0009-0006-8875-983X}
    \affiliation{%
      \institution{University of California, Irvine}
      \city{Irvine}
      \country{USA}
    }
    \email{shilonl2@uci.edu}
    
    \author{Changnam Hong}
    \orcid{0009-0000-3704-3379}
    \affiliation{%
      \institution{University of California, Irvine}
      \city{Irvine}
      \country{USA}
    }
    \email{changnah@uci.edu}

    \author{Qi Alfred Chen}
    \orcid{0000-0003-0316-9285}
    \affiliation{%
      \institution{University of California, Irvine}
      \city{Irvine}
      \country{USA}
    }
    \email{alfchen@uci.edu}
    
    \author{Joshua Garcia}
    \orcid{0000-0002-1696-8783}
    \affiliation{%
      \institution{University of California, Irvine}
      \city{Irvine}
      \country{USA}
    }
    \email{joshug4@uci.edu}

    \renewcommand{\shorttitle}{A Comprehensive Study of Bug-Fix Patterns in Autonomous Driving Systems}
    \renewcommand{\shortauthors}{Yuntianyi, Yuqi, Yirui, Shilong, Changnam, Alfred, and Joshua}

    \begin{abstract}

As autonomous driving systems (ADSes) become increasingly complex and integral to daily life, the importance of understanding the nature and mitigation of software bugs in these systems has grown correspondingly. 
\yuntianc{Addressing the challenges of software maintenance in autonomous driving systems (e.g., handling real-time system decisions and ensuring safety-critical reliability) is crucial due to the unique combination of real-time decision-making requirements and the high stakes of operational failures in ADSes.}
The potential of automated tools in this domain is promising, yet there remains a gap in our comprehension of the challenges faced and the strategies employed during manual debugging and repair of such systems.
In this paper, we present an empirical study that investigates \bfps in ADSes, with the aim of improving reliability and safety. We have analyzed the commit histories and bug reports of two major autonomous driving projects, Apollo and Autoware, from \numbug bug fixes with the study of bug symptoms, root causes, and \bfps.
Our study reveals several dominant \bfps, including those related to path planning, data flow, and configuration management. Additionally, we find that the frequency distribution of \bfps varies significantly depending on their nature and types and that certain categories of bugs are recurrent and more challenging to exterminate. 
Based on our findings, we propose a hierarchy of ADS bugs and two taxonomies of \numsyn syntactic \bfps and \numsem semantic \bfps that offer \yuntianyihl{guidance} for bug identification and resolution. 
\yuntianyihl{We also contribute a benchmark of \numbug ADS bug-fix instances.}

\end{abstract}

\ccsdesc[300]{Software and its engineering~Software libraries and repositories}
\ccsdesc[300]{Software and its engineering~Software defect analysis}

\keywords{Bug-fix pattern, Autonomous driving systems, Empirical study}
    
    \maketitle

    \blfootnote{Manuscript accepted by FSE 2025}

    \section{Introduction}\label{sec:introduction}

The rise of Autonomous Vehicles (AVs) signifies a crucial transformation in the realm of transportation. With more than 50 companies such as Ford, Toyota, Tesla, and Waymo~\cite{waymo,av_ford,av_intel,40_plus_corporations,18_av_companies,tesla_sold_2_million_cars,toyota_av,gm-av_charging} actively developing AVs, these vehicles are swiftly becoming a significant part of our daily lives.
This surge in development and deployment underscores the importance of understanding and maintaining the software that powers these complex cyber-physical systems. 
In the intricate ecosystem of AV software development, a critical yet often overlooked aspect is the pattern of bug fixes. While the initial focus of AV development has been on achieving functional and Operational Design Domain (ODD)~\cite{czarnecki2018operational} goal which specifies the operating conditions under which an ADS can operate safely, it is equally important to understand and repair the software bugs that inevitably arise. 
\yuntianyihl{We define a \bfp (BFP) as a recurring repair solution applied to similar bug types, capturing the modifications needed to resolve the bug.}
The significance of \bfps lies in their ability to reveal underlying system vulnerabilities and inform more effective automated repair strategies. This understanding is not only critical for enhancing software reliability and performance but also crucial for ensuring the safety of AV operations. As AVs find their way onto public roads, the stakes of software malfunctions escalate, necessitating a thorough and detailed study of bug repair.

One of the critical gaps in previous research~\cite{PanKW09,ZhongS15,SotoTWGL16,CamposM17,IslamZ20} on \bfps is the limited focus on syntactic \bfps, primarily at the code or statement level, often overlooking the semantic aspects (e.g., affected algorithms or domain-specific components) of the projects under study.
Although some existing empirical studies have investigated \bfps or strategies in a variety of domains including the deep learning stack~\cite{Huang0WCM023}, federated learning systems~\cite{DuCC0C023}, and deep learning libraries~\cite{IslamPNR20}, none of them have focused on \bfps in the ADS domain.

Our research centers on two widely-used~\cite{udacity_apollo,baidu_volvo_ford}, production-grade (i.e., used by some commercial companies) ADSes---Baidu Apollo~\cite{baidu_apollo}, which have reached mass production agreements with Ford and Volvo~\cite{baidu_volvo_ford}, and Autoware~\cite{autoware}, which was selected by the US Department of Transportation for intelligent transportation solutions~\cite{carma_github}. 
These systems represent the forefront of ADS technology and offer a large-scale dataset of bug history for understanding \bfps in real-world, high-stakes, or virtual-simulation environments. The exploration of BFPs in these systems is crucial, as it provides insights into the challenges and strategies employed in maintaining software that directly impacts the safety and reliability of AVs.
The importance of this study is further highlighted by the safety-critical nature of autonomous driving. As AVs continue to integrate into the fabric of daily transportation, the implications of software faults range from minor malfunctions to catastrophic failures, highlighting the necessity of meticulous software maintenance and bug fixing. While automated tools for bug detection and repair show promise, a comprehensive understanding of BFPs, encompassing both syntactic and semantic aspects, is crucial to advance the ADS domain.

In addressing this need, our study not only investigates the common syntactic BFPs in Apollo and Autoware but also delves into the semantic information underlying these patterns. This approach allows for a more nuanced understanding of the bugs, their causes, and the most effective strategies for their repair. 
We further introduce the concept of modularization granularity as a structured hierarchy of ADS bugs at various levels of abstraction. This hierarchy is essential for the systematic analysis and debugging of ADS and serves as a foundational structure for developing robust software systems. From both syntactic and semantic levels, our study formulates a more effective response to the complexities of ADS development and bug repair, potentially leading to resilient and fault-tolerant system designs.
Our proposed hierarchical classification of ADS \bfp study includes \numsyn syntactic and \numsem semantic \bfps, as well as root causes, symptoms, modules, sub-modules, related algorithms, \yuntianyihl{and detailed \bfas}. This taxonomy is instrumental for developers, testers, and researchers in developing automated bug detection and repair tools for ADS.
\yuntianc{Bug-fix patterns are the building blocks for repairing programs.}
Our taxonomy and statistics provide an in-depth understanding of common BFPs in ADSes, which ensures comprehensive coverage and paves the way for targeted improvements and innovations in debugging or repair techniques tailored for ADS.
\yuntianc{Specifically, our taxonomy can inform the design of new testing techniques, such as mutation testing, and the benchmark we propose can serve as a valuable resource for failure prediction research.}
The methodology of this study involves a detailed analysis of the collected data to identify common BFPs and understand their underlying causes and symptoms. 
Our research advances software engineering for ADS in the following key aspects:
\begin{itemize}[leftmargin=*,nosep]
\item We present the first empirical study of \bfps in the ADS domain through an analysis of \numbug fixed bugs from 15,099 pull requests in two open-source ADSes, offering a foundational understanding of the challenges and strategies in this domain. 

\item We \yuntianyihl{conduct the first study that} introduces and differentiates the concepts of syntactic and semantic \bfps, which serve as a practical guide for ADS software maintenance and identify opportunities for improved bug identification and repair for future work. 

\yuntianyihl{\item We propose a hierarchy of ADS \bfp study classified by modularization granularity. We also contribute a benchmark of \numbug ADS bug-fix instances, which is publicly accessible~\cite{bfp_artifacts}.} 
\end{itemize}

    \section{Background}\label{sec:background}

\subsection{Autonomous Driving Systems}

An ADS aims to achieve high automation levels for vehicles to automatically run on roads. 
The Society of Automotive Engineers defines six levels of autonomous driving, from Level 0, with no assistance systems, to Level 5, which represents fully autonomous driving \cite{RodelSMT14}. 
Baidu Apollo \cite{apollo} and Autoware~\cite{autoware} achieve Level 4~\cite{apollo_level_4}, which refers to a high degree of automation where the vehicle can handle all aspects of driving in certain environments without human intervention, but a human override is still an option.
The core modules and components are listed in \autoref{tab:modules_ads}.

\begin{table}[ht]
    \centering
    \caption{Modules in Autonomous Driving Systems}
    \begin{adjustbox}{width=0.9\linewidth}
    \begin{tabular}{|l|l|}
    \hline
    \textbf{Module} & \textbf{Description} \\
    \hline
    Planning & Makes decisions for the AV to execute, such as cruising or stopping. \\
    \hline
    Perception & Processes data from the surrounding environment detected by sensors. \\
    \hline
    Prediction & Receives obstacle information and predicts its future motion. \\
    \hline
    Control & Enforces the planned trajectory with lateral and longitudinal control. \\
    \hline
    Localization & Provides location, heading, and velocity information of the AV. \\
    \hline
    Simulator & Tests the ADS in a virtual environment, replicating real-world scenarios. \\
    \hline
    Sensing & Detects obstacles and traffic to understand the environment. \\
    \hline
    CAN Bus & Handles communication between software and vehicle. \\
    \hline
    HMI & Collects and visualizes the status and interfaces of the system. \\
    \hline
    HD Map & Includes lane geometries and locations of traffic control devices. \\
    \hline
    System & Coordinates the integration and operation of all AV modules. \\
    \hline
    Infrastructure & Includes robotics middleware to support communication among modules. \\
    \hline
    Utilities\&Tools & Provides necessary utilities, such as sensor calibration tools. \\
    \hline
    Docker & Includes the Docker image housing an instance of the ADS. \\
    \hline
    Build & Compiles and integrates the AV software, enabling smooth deployment. \\
    \hline
    Drivers & Contains the hardware drivers necessary for operating the AV. \\
    \hline
    Documentation & Provides guides, specifications, and manuals detailing the ADS. \\
    \hline
    \end{tabular}
    \end{adjustbox}
    \label{tab:modules_ads}
\end{table}

\subsection{Bug Study of Autonomous Driving Systems}

Understanding the nature of defects in ADSes is essential for enhancing their safety and performance. A comprehensive analysis of prior work \cite{GarciaF0AXC20} provides a foundational understanding by categorizing ADS bugs from the root causes that reflect errors in code and symptoms manifested as incorrect behaviors or errors during runtime as presented in \autoref{tab:root_causes_ads} and \autoref{tab:symptoms_ads}.

\vspace{-1.5ex}
\begin{table}[ht!]
    \centering
    \begin{minipage}{0.52\linewidth}
        \centering
        \caption{Root Causes of Bugs in the ADS}
        \begin{adjustbox}{width=\linewidth}
        \begin{tabular}{|l|l|}
        \hline
        \textbf{Type} & \textbf{Description} \\
        \hline
        Alg & Flawed logic that requires a comprehensive correction. \\
        \hline
        Num & Incorrect numerical calculations, values, or usage. \\
        \hline
        Assi & Wrong variable assignments or initializations. \\
        \hline
        MCC & Absence of necessary conditional statements. \\
        \hline
        Data & Incorrect data structure definitions or pointer misuse. \\
        \hline
        Exter-API & Incorrect usage of interfaces of other systems or libraries. \\
        \hline
        Inter-API & Misuse of interfaces of other components within the ADS. \\
        \hline
        ICL & Incorrect condition logic or faulty conditional expressions. \\
        \hline
        Conc & Mismanagement of concurrency mechanisms like threads. \\
        \hline
        Mem & Misuse of memory or improper memory management. \\
        \hline
        Doc & Incorrect manuals, tutorials, or code comments. \\
        \hline
        Config & Incorrect system setup or build settings in configurations. \\
        \hline
        OT & Infrequent and uncategorized issues. \\
        \hline
        \end{tabular}
        \end{adjustbox}
        \label{tab:root_causes_ads}
    \end{minipage}
    \hfill
    \begin{minipage}{0.47\linewidth}
        \centering
        \caption{Symptoms of Bugs in the ADS}
        \begin{adjustbox}{width=\linewidth}
        \begin{tabular}{|l|l|}
        \hline
        \textbf{Type} & \textbf{Description} \\
        \hline
        Crashes & Improper termination of an ADS. \\
        \hline
        Hangs & The system becomes unresponsive but remains running. \\
        \hline
        Build & Errors hinder the compilation or installation of ADS modules. \\
        \hline
        DGUI & Erroneous output on a GUI, visualization, or HMI. \\
        \hline
        Cam & Errors prevent image capture by the camera. \\
        \hline
        Stop & Incorrect behaviors during stopping or parking. \\
        \hline
        LPN & Incorrect lane positioning, maintenance, or navigation. \\
        \hline
        SVC & Incorrect AV speed and velocity management. \\
        \hline
        TLP & Incorrect behaviors involving handling of traffic lights. \\
        \hline
        Lau & ADS or component fails to start or launch. \\
        \hline
        Turn & AV behaves incorrectly when making a turn. \\
        \hline
        Traj & Incorrect trajectory prediction results. \\
        \hline
        IO & Incorrect input/output operations to files or devices. \\
        \hline
        LOC & Inaccurate multi-sensor fusion-based localization. \\
        \hline
        SS & Behaviors that affect the safety or security of its passengers. \\
        \hline
        OP & Incorrect processing or handling of obstacles on the road. \\
        \hline
        Logic & Incorrect behaviors that do not terminate the program. \\
        \hline
        Doc & Errors in the documentation or comments. \\
        \hline
        UN & Symptoms that are unreported and cannot be identified. \\
        \hline
        OT & Infrequent symptoms that do not fit into above categories. \\
        \hline
        \end{tabular}
        \end{adjustbox}
        \label{tab:symptoms_ads}
    \end{minipage}
\end{table}

\subsection{\BFPs}

The concept of \bfps (BFPs) emerges as a critical tool in the software engineering domain, offering a structured approach to diagnosing and rectifying common defects in software systems~\cite{PanKW09,CamposM17}. These patterns, distilled from the collective experience of developers, encapsulate proven solutions to recurring problems, thereby streamlining the debugging and maintenance processes. In the context of ADS, where software reliability and safety are essential, the identification and application of BFPs become even more vital.
Autonomous driving systems, with their integration of complex algorithms, sensor data processing, and real-time decision-making, present a unique set of challenges for software maintenance. The diversity of bugs in ADS---ranging from algorithmic inaccuracies and sensor data misinterpretations to issues in system integration and performance---necessitates a specialized set of BFPs tailored to these specific problems. For instance, a \bfp aimed at addressing inaccuracies in environmental perception might involve strategies for sensor data fusion enhancement or algorithmic refinement for object detection.

    \section{Methodology}\label{sec:methodology}

\subsection{Dataset}

In our study of \bfps, we collected a dataset that encompasses 19,379 commits, 9,708 merged pull requests, and 3,532 issues for Apollo~\cite{apollo}, as well as 22,825 commits, 5,391 merged pull requests, and 559 issues for Autoware. 
All these commits, issues, and pull requests were established on or prior to November 13, 2023, as detailed in \autoref{tab:dataset}.

\begin{wraptable}[7]{r}{0.52\textwidth}
\vspace{-1ex}
\centering
\caption{Statistics of ADS Dataset}
\begin{adjustbox}{width=\linewidth}
\begin{tabular}{ccccccc}
\Xhline{3\arrayrulewidth}
    \textbf{System} & \textbf{\makecell{Start Date--\\End Date}} & \textbf{Commits} & \textbf{Issues} & \textbf{\makecell{Pull\\Requests}} & \textbf{\makecell{Bug\\Fixes}} \\
\hline
Apollo & \makecell{07/04/2017--\\11/13/2023} & 19,379 & 3,532 & 9,708 & 338 \\
\hline
Autoware & \makecell{08/25/2015--\\11/13/2023} & 22,825 & 559 & 5,391 & 993 \\
\Xhline{3\arrayrulewidth}
\end{tabular}
\end{adjustbox}
\label{tab:dataset}
\end{wraptable}

Considering the objective to identify \bfps in ADSes, we collected closed and merged single-file pull requests that fix bugs. Such pull requests enable us to 
(1) scrutinize the modified source code, associated issues, and developer discussions; 
and (2) avoid too many bugs and \bfps that occurred in one pull request to interfere with the evaluation and analysis. 
Note that GitHub's pull requests serve diverse functions, such as introducing new features, bug fixes, or refactoring. 
We implemented a technique to maximize the capture of bug-fix pull requests. In doing so, we employed an approach similar to those of previous research~\cite{GarciaF0AXC20,IslamNPR19,VasilescuYWDF15,ZhangCCXZ18} by curating a list of bug-associated keywords like ``fix'', ``repair'', ``defect'', ``error'', ``bug'', ``issue'', ``mistake'', ``incorrect'', ``fault'', and ``flaw'', subsequently searching for these terms in both titles and content.
Pull requests with any of these keywords were classified as initial bug-fix requests. We further filtered out 62 pull requests that were not real bug fixes even with keywords and repeated pull requests derived from different branches during labeling by analyzing their description and intentions.
This methodology yielded 338 and 993 integrated pull requests for Apollo and Autoware, respectively, that fit our criteria.

\subsection{Hierarchy of \yuntianyihl{ADS \BFP Study by Modularization Granularity}}

In this study, we aim to analyze the nature of ADS \bfps by examining them through seven distinct perspectives: (i) the root causes indicative of developer errors in the codebase; (ii) the symptoms of these bugs, as evidenced by aberrant behaviors, system failures, or runtime errors; (iii) the specific ADS module and sub-module where the bug is localized; (iv) the algorithm or function that the bug is related to; (v) the syntactic \bfp; (vi) the semantic \bfp \yuntianyihl{; and (vii) the \bfa related to the specific \bfp, root cause, sub-module, or algorithm}.

\begin{figure}[ht]
\centering
\includegraphics[width=\linewidth]{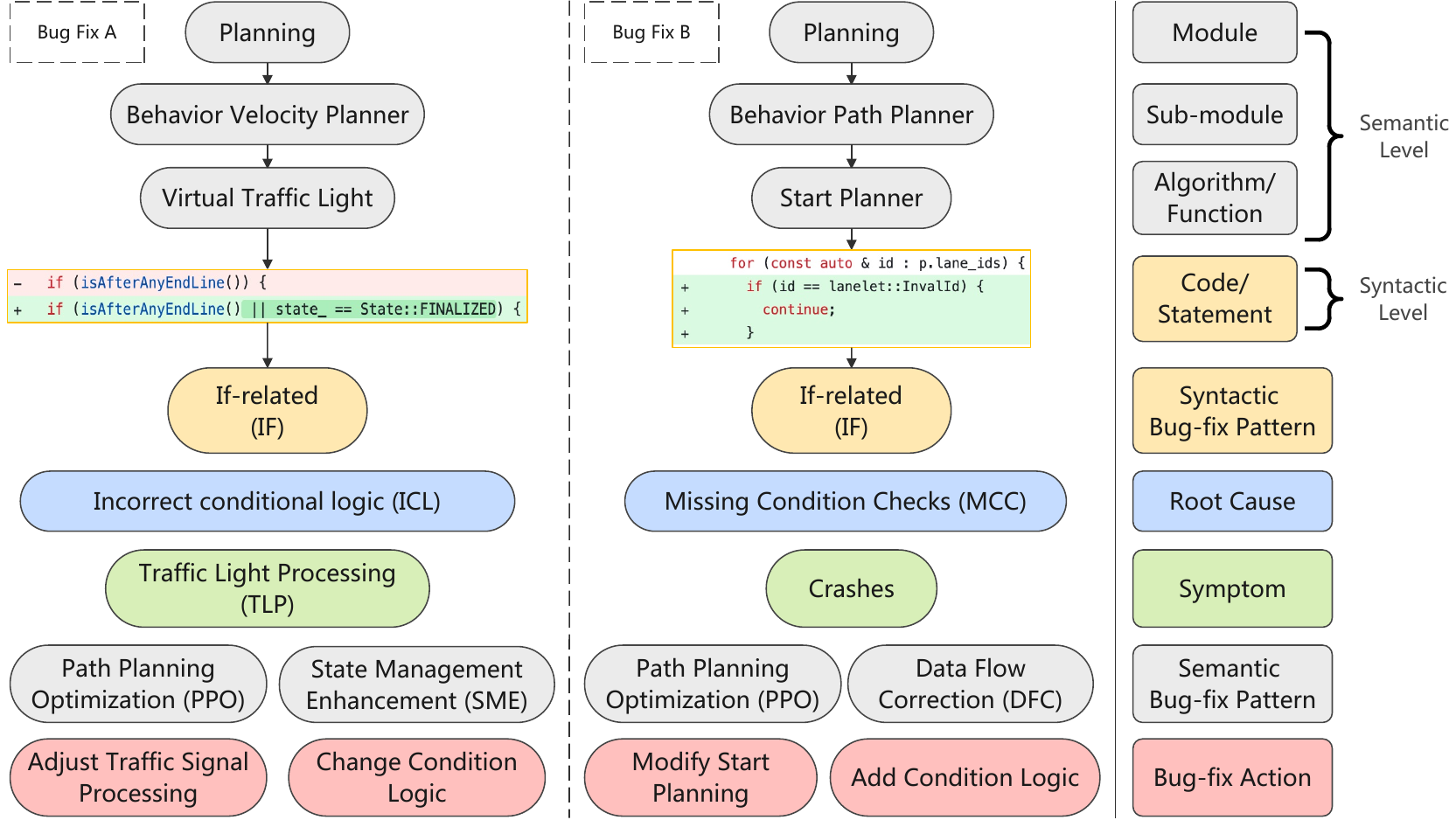}
\caption{Hierarchy of ADS \yuntianyihl{\BFP by Modularization Granularity}}
\label{fig:hierarchy}
\end{figure}

We introduce the concept of \textit{Modularization Granularity} as a means to dissect and understand the architecture of ADS, offering a clear depiction of the structural layers of software development into a hierarchy of granularity that spans from broad architectural modules to the minute intricacies of code statements.
Modularization granularity is essential to decompose complex systems into manageable segments. We identify this granularity through various levels of labels, each serving its unique role in the modular architecture.
\autoref{fig:hierarchy} shows the hierarchy along with two examples presenting how we label the bug fixes in the Planning module.

On the semantic level, we have \textit{Module} and \textit{Sub-module} labels which indicate the major components and their subdivisions respectively, facilitating a high-level understanding and management of the system. We also have \textit{Algorithm/Function} labels that reveal the intricate details of the system's functionality and operational logic. Finally, at the most granular level, we have the \textit{Statement/Code} labels, which are indicative of the syntactic construction of the software.
\textit{Syntactic BFPs} are categorized according to \yuntianyihl{code and} statement changes, while \textit{Semantic BFPs} \yuntianyihl{are analyzed from a high-level comprehensive understanding of ADS misbehaviors}.
\yuntianyihl{\textit{\Bfas} are determined through an in-depth examination of other labels, which describe the actionable strategies made to bugs, providing detailed steps and modifications necessary to correct errors, adjust code behavior, and improve system functionality.
For each pull request, we also label the \textit{Symptom} to identify the observable effects of bugs and the \textit{Root cause} that delves into their underlying reasons to help understand the relations between \bfps and bug types. This hierarchical framework allows us to trace symptoms back to their root causes and categorize them into our taxonomy of \bfps. This structured analysis informs our findings by providing a nuanced view of how different types of bugs arise, propagate through, and influence the system.}

While distinguishing the severity of individual bug-fix patterns may be challenging, the design of our modularization granularity hierarchy allows us to use the symptoms to assess severity. For instance, the bugs that have Crashes symptoms are clearly more severe than those with DGUI or no symptoms. We could use symptoms and corresponding violations to assess the severity of bugs. 

\yuntianc{The hierarchy has practical applications for both researchers and practitioners. For researchers, it provides a robust framework for studying bug-fix patterns in other domains. Its top three semantic levels and semantic bug-fix patterns can be customized to suit specific application domains, making it broadly applicable. For example, adapting the hierarchy more broadly to cyber-physical systems or robotics could provide valuable insights into their unique bug-fix challenges. This flexibility underscores the hierarchy's potential to standardize bug-fix analysis across diverse systems, aiding researchers in labeling and analyzing bug-fix data systematically. For practitioners, the hierarchy serves as a diagnostic tool to guide debugging and repair processes. By mapping the symptom and root cause of a bug to specific semantic or syntactic bug-fix patterns, the hierarchy helps practitioners identify actionable bug-fix actions.
Furthermore, it connects these patterns to affected modules, sub-modules, algorithms, and code statements, providing targeted guidance for debugging and repair. The hierarchy's adaptability also ensures its applicability to other modular architectures. 
}
\subsection{Classification and Labeling}

To mitigate potential subjective biases during the labeling phase, we allocated each of the \numbug merged bug-fix pull requests to two co-authors, who are also contributors and developers of both Apollo and Autoware open-source projects and possess enough background in the ADS domain. 
We employed the open-coding strategy of intercoder reliability~\cite{intercoder_reliability} to help strengthen the labeling process.
Our methodology necessitated that each co-author independently assess the bug, which involved meticulous examination of the source code, commit logs, code reviews, pull request information, and associated issue descriptions to recognize the labeling items.

In our work, we commenced with the root cause and symptom taxonomies in a previous ADS bug study~\cite{GarciaF0AXC20} and also taxonomy of generic \bfps focusing on the syntactic level~\cite{PanKW09,SotoTWGL16,CamposM17,IslamZ20}
as a foundation for ADS bug analysis. 
The taxonomy of root causes was subsequently augmented by employing an open-coding paradigm, thereby broadening the spectrum. 
For the pull request that eluded classification within the foundational taxonomy, each co-author designated a label for it. Post-labeling, co-authors collaboratively reconciled any discrepancies in their classifications. 
We added two root causes, \textbf{Syntax, Naming, and Typography (SNT)} that involves errors in the basic structure of the code, including syntactical mistakes, naming conventions, and typographical errors, and \textbf{Dependency Issues (DEP)} related to importing, versioning, and managing dependencies.

For semantic \bfps ~\yuntianyihl{and \bfas}, none of the previous work could provide a useful taxonomy due to the domain-specific nature of ADS. In this research, the open coding process was employed to refine and identify distinct semantic \bfps~\yuntianyihl{and \bfas}.
A preliminary investigation was conducted involving two co-authors who independently examined the bug fixes to establish a tentative classification framework. Each rater recommended a series of categories, which were later amalgamated and refined during a face-to-face session attended by all contributing authors. This meeting served as a platform to validate and integrate the classification schemes proposed by the individual raters. This reconciliation process led to updates in the classification scheme. 
\yuntianyihl{We validated the final classification by consulting with researchers in the ADS domain and developers from Apollo and Autoware. Their expertise helped refine the classification scheme, and we relabeled the affected pull requests accordingly.}
The classification result is presented in \autoref{sec:taxonomy}.
Our findings indicated that a single bug origin may result in multiple symptoms and \bfps. Therefore, certain bugs were cataloged into multiple categories, unlike the previous study~\cite{GarciaF0AXC20} only considered a single symptom for each bug.
In the process of labeling, two co-authors, possessing expertise in ADS, were engaged to categorize the \bfps according to the specified schema. 
The degree of concordance between these raters was quantified using Cohen's Kappa coefficient~\cite{VieiraKS10}, which was used by a recent \bfp study~\cite{IslamPNR20}. 
In instances of labeling discrepancies, periodic discussions were conducted to achieve reconciliation. 
Throughout this process, the Kappa score persistently exceeded 80\%, indicative of a robust understanding and unanimous agreement among the raters~\cite{mchugh2012interrater}.

    \section{Taxonomy}\label{sec:taxonomy}


\subsection{\BFPs at the Syntactic Level}
\label{subsec:syntactic_taxonomy}

We adopted the classification from previous work~\cite{PanKW09,IslamZ20} as it comprehensively covered the syntactic BFPs observed in Apollo and Autoware. We also added new patterns like \yuntianyihl{Syntax (SYN),} \textit{Document Fix (DOC)}, and \textit{Library (LIB)} to describe bug repairs that are missing in previous studies but relevant for ADSes. We enumerate the syntactic \bfps in the following paragraphs.

\patternpara{If-related (IF)}
Modifications in the conditional logic of code through addition, modification, or removal of \textit{if} statements to ensure that operations are executed under the correct conditions.

\patternpara{Assignment (AS)}
Modifications in the assignment statements, including changes in the expressions on the right-hand side of an assignment, to ensure correct value assignment and operation.

\patternpara{Method Call (MC)}
The MC pattern involves adjustments in the way methods are invoked, which include changing the number and type of parameters or the values passed to it.

\patternpara{Method Declaration (MD)}
Changes in Method Declarations involve alterations in the method's interface, such as the number of parameters, parameter types, or return type. 

\patternpara{Sequence (SQ)}
Changes in the order of a series of operations or method calls. 

\patternpara{Loop (LOOP)}
Modifications in the conditions or the structure of loop statements. 

\patternpara{Return (RT)}
This pattern encompasses modifications in the return statements of functions, ensuring that functions return the correct values or types as expected.

\patternpara{Local Variable (LV)}
Changes to the declaration or initialization of local variables within functions. 

\patternpara{Non-source-code Variable (NV)} 
NV adjustments pertain to changes in parameters that are not part of the source code but are essential for the configuration of the software.

\patternpara{Library (LIB)}
Fixes involve the addition, removal, or modification of libraries used in the project. 

\patternpara{Syntax (SYN)}
This pattern addresses issues like incorrect indentation \yuntianyihl{in Python}, syntax misuse, or typographical errors \yuntianyihl{that can affect the code execution}.

\patternpara{Try-Catch (TRY)}
This pattern involves the addition or removal of try-catch blocks, which are vital for error handling and exception management without crashing.

\patternpara{Switch (SW)}
Add or remove switch branches or change the conditions within the switch block. 

\patternpara{Class Field (CF)}
CF fixes pertain to modifications in the class fields or attributes that are present in any object of a class, and whose lifetime is the same as the object lifetime. 

\patternpara{Document Fix (DOC)}
Corrections or updates in documentation or comments within the code. 

\subsection{\BFPs at the Semantic Level}
\label{subsec:semantic_taxonomy}

The semantic level of \bfps in ADS delves into the deeper, meaning-oriented aspects of software bugs, focusing on how these bugs affect the behavior and functionality of the system as a whole. Unlike syntactic \bfps, which are concerned with the structure and form of the code, semantic \bfps address the logic, algorithms, and operational semantics of the ADS software, sometimes including multiple syntactic patterns within a single semantic pattern. To comprehensively understand these patterns, we categorize them into \textit{domain-specific} and \textit{domain-independent} \bfps. Domain-specific patterns address issues unique to autonomous driving technology, while domain-independent patterns apply to a broader range of software systems. 
We enumerate our identified ADS semantic \bfps in the remainder of this section.

\subsubsection{Domain-Specific Semantic \BFPs.}

\yuntianyihl{Domain-specific patterns are uniquely tailored to address the problems inherent to ADSes. 
These patterns often highlight the intricate interactions between sub-modules and external factors like road conditions, traffic regulations, and sensor data.}

\patternpara{Path Planning Optimization (PPO)}
Refine the algorithms related to path and velocity planning in ADS, including enhancements to the lane-change strategies, obstacle avoidance, or navigation in complex environments.
The code snippet provided illustrates a specific bug fix of PPO:

\begin{lstlisting}
+ overwriteStopPoint(clipped,traj_smoothed);
  traj_smoothed.insert(traj_smoothed.begin(),traj_resampled->begin(),traj_resampled->begin()+ *traj_resampled_closest);
- overwriteStopPoint(*traj_resampled,traj_smoothed);
\end{lstlisting}

\noindent 
In this fix, the function \texttt{overwriteStopPoint} is being called with updated parameters while being moved forward.
By adjusting the source of the stop point data and how the trajectory is assembled, the updated code aims to improve the vehicle's path planning accuracy, making the stop points more precise and better aligned with the actual route and obstacles. 

\patternpara{Sensor Data Interpretation (SDI)}
Address issues in interpreting data from sensor input (e.g., \lidar, radar, and cameras), ensuring accurate and reliable environmental perception.

\patternpara{Control System Adjustment (CSA)}
Optimize vehicle's control systems, such as steering, braking, and throttle, which are vital to ensure smooth and safe vehicle operation.

\patternpara{Predictive Algorithm Enhancement (PAE)}
Improve the algorithms responsible for predicting potential hazards or traffic conditions for proactive safety measures and efficient route planning.

\patternpara{SLAM Algorithm Refinement (SAR)}
Refine Simultaneous Localization and Mapping (SLAM) algorithms used in the Localization module to enhance map creation and position estimation. 

\patternpara{Environmental Adaptability (EA)}
EA involves enhancements in the system's ability to adapt to different environmental conditions, such as varying weather, lighting, or road surface conditions, which are critical for the reliability of ADSes.
The code snippet below illustrates an EA bug fix:

\begin{lstlisting}
  if ((rclcpp::Time(msg->stamp)-rclcpp::Time(latest_perception_msg_.stamp)).seconds() > perception_time_tolerance_) {
-   latest_external_msg_.signals.clear();}
+   latest_perception_msg_.signals.clear();}
\end{lstlisting}

\noindent 
The repair suggests the original code was incorrectly resetting wrong signals when a delay in perception message updates was detected. By clearing correct signals, the system ensures outdated perception data does not interfere with the ability to adapt to current environmental conditions.

\patternpara{Firmware-Software Harmonization (FSH)}
Align the vehicle's firmware with higher-level software systems, ensuring that hardware-software interactions are seamless and efficient.

\patternpara{Module Integration and Interaction (MII)}
Improve the integration and interaction among various ADS modules for ensuring cohesive and harmonious system operations.

\patternpara{Safety Protocol Enhancement (SPE)}
Enhance safety features to ensure compliance with emerging safety standards and regulations, as well as to address newly identified safety concerns.

\patternpara{Communication Protocol Refinement (CPR)}
Enhance communications for vehicle-to-vehicle (V2V) and vehicle-to-everything (V2X) interactions to ensure robust and secure data transmission.

\patternpara{Real-time Data Processing Improvement (RDPI)}
Improvements in algorithms for ensuring efficient handling of real-time data across modules and sub-modules for better vehicle control.

\patternpara{Autonomous Decision-Making Improvement (ADMI)}
Enhancements in ADS decision-making algorithms, ensuring that the vehicle makes safe and logical decisions in real-time traffic scenarios.

\patternpara{Simulation Performance Optimization (SPO)}
Improve the simulator's performance to enable more efficient virtual testing. This pattern could involve code optimizations to reduce computational load, enhance rendering speeds, and decrease latency in the simulation environment.

\subsubsection{Domain-Independent Semantic \BFPs.}

Domain-independent semantic \bfps play a vital role in the overall functionality and robustness of ADSes. These patterns are not unique to ADS but are critical for ensuring the software's operational efficacy and reliability.

\patternpara{Data Flow Correction (DFC)}
Address issues in the way data is passed and used throughout the system.
Rectify the data flow between functions to ensure data is correctly processed and utilized.

\patternpara{Logic Amendment (LA)}
Changes in the underlying logic of the code, which is applied when existing logic leads to incorrect or undesired behaviors to realign with the intended functionality.

\patternpara{State Management Enhancement (SME)}
Refine how states (e.g., vehicle state or sensor state) are tracked, updated, and managed within the system to ensure correct state-dependent functionalities.

\patternpara{Interface Consistency (IC)}
Interface consistency fixes are applied to standardize and streamline the interactions between different software modules or components. 

\patternpara{Resource Management Improvement (RMI)}
Optimize the utilization of system resources like memory, processing power, and network bandwidth that can impact system performance.

\patternpara{Concurrency Control (CC)}
Address issues arising from simultaneous operations or multi-threaded environments, ensuring concurrent processes do not lead to conflicts, deadlocks, or inconsistency.

\patternpara{Error Handling Refinement (EHR)}
Improve the system's ability to detect, report, and recover from errors. Robust error handling can prevent cascading failures and improve system resilience.

\patternpara{Configuration and Environment Management (CEM)}
Update configuration or environment options to ensure that the ADS operates correctly and is adaptable to different operational contexts.

\patternpara{Security Strengthening (SS)}
Enhance the system's defense against external threats and vulnerabilities, including security flaws, which is important in ADS due to its autonomous nature.

\patternpara{User Interface Adjustment (UIA)}
Changes to the human-machine interface of ADS, aimed at improving usability, providing clearer information, or enhancing the overall user experience.

\patternpara{Dependency Update (DU)}
Update or modify external libraries or components to ensure compatibility and that the system benefits from the latest improvements of its dependencies.

\patternpara{Build and Compilation Enhancement (BCE)}
This pattern aims to improve the efficiency, reliability, and accuracy of the build and compilation processes in software development.

\patternpara{Debugging Tools Improvement (DTI)}
DTI involves integrating better logging systems, visual debugging aids, and automated analysis tools to help identify and diagnose issues.

\patternpara{Documentation Update and Clarification (DUC)}
This pattern involves correcting and updating the documentation to ensure that it reflects the current state of the project accurately. 

    \section{Experimental Results}\label{sec:evaluation}

This paper investigates \bfps in ADSes to address the following research questions:

\vspace{-1.5mm}
\framed
    \vspace{-2mm}
    \noindent
    \textbf{RQ1 (\Bfp frequency):} What are the most common \bfps in the ADS?
    \vspace{-2mm}
\endframed
\vspace{-1.5mm}

\yuntianyihl{This question aims to unravel the prevalent methodologies employed to address syntactic bugs—those related to the structure or format of the code—and semantic bugs, which concern the meaning or behavior of the code within ADS.}

\vspace{-1.5mm}
\framed
    \vspace{-2mm}
    \noindent
    \textbf{RQ2 (\Bfa frequency):} What are the most common \bfas in the ADS?
    \vspace{-2mm}
\endframed
\vspace{-1.5mm}

This question investigates the frequency of specific \bfas and strategies in the ADS. By categorizing these actions into domain-specific and domain-independent fixes, the study aims to provide actionable insights into which strategies are most frequent in bug fixing.

\vspace{-1.5mm}
\framed
    \vspace{-2mm}
    \noindent
    \textbf{RQ3 (\Bfps across root causes):} How do semantic \bfps vary across different root causes of bugs in the ADS?
    \vspace{-2mm}
\endframed
\vspace{-1.5mm}

\yuntianyihl{This analysis delves into the diversity of bug origins within ADS and examines the correspondence between these root causes and the \bfps applied. Understanding this correlation is important for tailoring debugging strategies to the specific nature of a bug’s root cause.}

\vspace{-1.5mm}
\framed
    \vspace{-2mm}
    \noindent
    \textbf{RQ4 (\Bfps across symptoms):} How do \yuntianyihl{semantic} \bfps vary across different symptoms in the ADS?
    \vspace{-2mm}
\endframed
\vspace{-1.5mm}

\yuntianyihl{This question focuses on the manifestation of bugs through various symptoms within ADS.
By mapping \bfps to specific symptoms, the study aims to elucidate targeted strategies for symptom-based debugging, fostering symptom-oriented guidance to ADS optimization.}
    
\vspace{-1.5mm}
\framed
    \vspace{-2mm}
    \noindent
    \textbf{RQ5 (\Bfps across ADS modules):} Are \yuntianyihl{semantic} \bfps different across various modules in the ADS? 
    \vspace{-2mm}
\endframed
\vspace{-1.5mm}

\yuntianyihl{ADS encompasses a spectrum of modules, 
this question explores whether the \bfps employed are module-specific, reflecting the unique challenges and requirements of each module.}

\subsection{\textbf{RQ1: \BFP Frequency}}

For RQ1, we analyzed the frequency and types of common \bfps in autonomous driving systems, specifically focusing on Apollo and Autoware. The results, presented in \autoref{tab:syntactic_distribution_bfp} and \autoref{tab:semantic_distribution_bfp}, categorize the \bfps into syntactic and semantic types.

\begin{wraptable}[13]{r}{0.45\textwidth}
\centering
\vspace{-2ex}
\caption{Syntactic \BFPs in ADSes}
\vspace{-0.5ex}
\label{tab:syntactic_distribution_bfp}
\begin{adjustbox}{width=\linewidth}
\begin{tabular}{l|c|c|c}
\toprule
\textbf{Syntactic \BFP} & \textbf{Apollo} & \textbf{Autoware} & \textbf{Both} \\
\midrule
If-related (IF) & 13.6\% & 19.5\% & 18.3\% \\
Assignment (AS) & 15.6\% & 14.7\% & 14.8\% \\
Method Call (MC) & 26.9\% & 23.1\% & 23.9\% \\
Method Declaration (MD) & 1.8\% & 1.9\% & 1.9\% \\
Sequence (SQ) & 2.8\% & 2.4\% & 2.5\% \\
Loop (LOOP) & 3.8\% & 6.0\% & 5.5\% \\
Return (RT) & 4.0\% & 8.4\% & 7.6\% \\
Local Variable (LV) & 9.5\% & 12.6\% & 12.0\% \\
Non-source-code Variable (NV) & 7.3\% & 5.9\% & 6.1\% \\
Library (LIB) & 2.8\% & 1.8\% & 2.0\% \\
Syntax (SYN) & 1.8\% & 0.1\% & 0.4\% \\
Try-Catch (TRY) & 0.6\% & 0.2\% & 0.3\% \\
Switch (SW) & 0.2\% & 0.1\% & 0.1\% \\
Class Field (CF) & 0.0\% & 0.1\% & 0.1\% \\
Documentation (DOC) & 9.1\% & 3.2\% & 4.4\% \\
\bottomrule
\end{tabular}
\end{adjustbox}
\end{wraptable}

In both ADSes, syntactic BFPs, particularly 23.9\% for Method Call (MC) and 18.3\% for If-related (IF), are most frequently observed. 
These patterns typically involve adjustments in the code's basic structure and flow, suggesting a recurring need for attention to fundamental coding practices in ADS development.
\yuntianyihl{The MC pattern represents nearly one-quarter of all bug-fix patterns in both Apollo (26.9\%) and Autoware (23.1\%). This suggests that method invocation errors are particularly challenging in ADSes. Given that method calls often involve integrating multiple components, it hints at the complexity and interdependencies within these systems. This could imply a need for better modularization or more robust testing around component interactions.}

\vspace{-1ex}
\begin{finding}
\label{finding:rq1_if_mc}
    Syntactic \bfps like 18.3\% of If-related (IF) and 23.9\% of Method Call (MC) are the most common types in ADSes, \yuntianyihl{which indicates that issues related to control flow decisions and incorrect method invocations are prevalent.}
\end{finding}
\vspace{-1ex}

Moreover, domain-specific semantic \bfps, including 22.9\% Path Planning Optimization (PPO) and 6.5\% Module Integration and Interaction (MII), also feature prominently in our findings. These patterns often involve complex modifications and a deep understanding of the system’s operational logic. 
The complexity and critical nature of these fixes, particularly in MII, reveal the intricacies involved in ADS software maintenance, highlighting the necessity for robust architectures. This complexity underscores the importance of both software researchers and practitioners in directing attention to the integration, modular architecture, and encapsulation of ADS~\cite{TasKZS16,0003LPZ0SG18}. Moreover, the significant presence of MII-related issues suggests an impending need to prioritize integration testing in ADS, an area that currently lacks comprehensive study but is crucial for vehicle safety and performance~\cite{LouDZZ022}.

\vspace{-1ex}
\begin{finding}
\label{finding:rq1_ppo_mii}
    Semantic \bfps like 22.9\% fixes of Path Planning Optimization (PPO) and 6.5\% of Module Integration and Interaction (MII) are the most common domain-specific ones. The prominence of MII-related fixes indicates the need for improved module integration testing techniques \yuntianyihl{and advanced modular ADS software architecture design and encapsulation}.
\end{finding}
\vspace{-1ex}

The domain-independent semantic patterns, such as 13.6\% of Data Flow Correction (DFC) 
are indicative of foundational challenges in ensuring cohesive system operation.
The prevalence of DFC in \bfps highlights a critical area of system functionality that often is understudied, pointing to a gap in data flow-oriented testing within the software testing domain~\cite{SuWMPHCS17,Weyuker90}. This gap suggests the need for more comprehensive data-flow testing strategies that move beyond traditional code coverage to enhance data-flow coverage and system reliability.

\vspace{-1ex}
\begin{finding}
\label{finding:rq1_dfc}
    The prominence of domain-independent semantic \bfps like 13.6\% of Data Flow Correction (DFC) indicates the need for more research on data flow-based testing  (e.g., data-flow coverage) for ADSes, which may be more computationally expensive but possibly worthwhile for these safety-critical ADSes.
\end{finding}
\vspace{-1ex}

\begin{wraptable}[22]{r}{0.58\textwidth}
\centering
\vspace{-2ex}
\caption{Semantic \BFPs in ADSes}
\vspace{-0.5ex}
\label{tab:semantic_distribution_bfp}
\begin{adjustbox}{width=\linewidth}
\begin{tabular}{l|c|c|c}
\toprule
\textbf{Semantic \BFP} & \textbf{Apollo} & \textbf{Autoware} & \textbf{Both} \\
\midrule
\textbf{Domain-specific Semantic Patterns} & & & \\
Path Planning Optimization (PPO) & 9.2\% & 27.3\% & 22.9\% \\
Sensor Data Interpretation (SDI) & 2.0\% & 2.4\% & 2.3\% \\
Control System Adjustment (CSA) & 1.2\% & 1.5\% & 1.4\% \\
Predictive Algorithm Enhancement (PAE) & 2.0\% & 0.6\% & 0.9\% \\
SLAM Algorithm Refinement (SAR) & 3.8\% & 2.9\% & 3.1\% \\
Environmental Adaptability (EA) & 4.7\% & 4.0\% & 4.2\% \\
Firmware-Software Harmonization (FSH) & 2.2\% & 0.4\% & 0.8\% \\
Module Integration and Interaction (MII) & 12.4\% & 4.6\% & 6.5\% \\
Safety Protocol Enhancement (SPE) & 0.0\% & 0.7\% & 0.5\% \\
Communication Protocol Refinement (CPR) & 0.5\% & 0.6\% & 0.6\% \\
Real-time Data Processing Improvement (RDPI) & 2.8\% & 0.7\% & 1.2\% \\
Autonomous Decision-Making Improvement (ADMI) & 1.3\% & 0.3\% & 0.5\% \\
Simulation Performance Optimization (SPO) & 0.2\% & 1.5\% & 1.1\% \\
\addlinespace
\textbf{Domain-independent Semantic Patterns} & & & \\
Data Flow Correction (DFC) & 11.2\% & 14.4\% & 13.6\% \\
Logic Amendment (LA) & 2.0\% & 2.1\% & 2.1\% \\
State Management Enhancement (SME) & 3.2\% & 6.1\% & 5.4\% \\
Interface Consistency (IC) & 0.3\% & 0.7\% & 0.6\% \\
Resource Management Improvement (RMI) & 1.2\% & 0.4\% & 0.6\% \\
Concurrency Control (CC) & 0.3\% & 0.3\% & 0.3\% \\
Error Handling Refinement (EHR) & 4.0\% & 3.3\% & 3.5\% \\
Configuration and Environment Management (CEM) & 7.2\% & 7.4\% & 7.3\% \\
Security Strengthening (SS) & 0.5\% & 0.1\% & 0.2\% \\
User Interface Adjustment (UIA) & 2.5\% & 2.4\% & 2.5\% \\
Dependency Update (DU) & 5.2\% & 3.0\% & 3.6\% \\
Build and Compilation Enhancement (BCE) & 8.5\% & 5.4\% & 6.2\% \\
Debugging Tools Improvement (DTI) & 4.0\% & 3.3\% & 3.5\% \\
Documentation Update and Clarification (DUC) & 7.5\% & 3.5\% & 4.5\% \\
\bottomrule
\end{tabular}
\end{adjustbox}
\end{wraptable}

\autoref{tab:semantic_distribution_bfp} \yuntianyihl{also indicates that Autoware has a higher ratio of Path Planning Optimization (PPO) \bfps, possibly due to its specific development focus and project needs since Autoware has more code lines and sub-modules in the Planning module than Apollo does. Autoware also claims to be a customizable and easily extendable Planning development platform~\cite{autoware_planning_design}. Its maintainers accepted more pull requests from community contributors around the world while Apollo's maintainers tended to merge more pull requests from Baidu employees.
Apollo's higher number of build, debug, and documentation-related instances reflect different project workflows and priorities, such as a stronger emphasis on ensuring robust documentation and debugging processes to facilitate collaboration and maintenance. This could be due to Apollo's larger user base and its need for clear documentation and thorough debugging to support a wide range of developers and applications. Additionally, Apollo's extensive integration with various hardware and software components, reflected by the Module Integration and Interaction (MII) patterns of 12.4\% in Apollo compared to 4.6\% in Autoware, may require more frequent build and debug fixes to maintain compatibility and performance. Autoware just finished its major version update to Autoware.universe~\cite{autoware_universe_transition}, the tutorial and documentation are still under construction and waiting to be completed in the future. 
}

\vspace{-1ex}
\begin{finding}
\label{finding:rq1_autoware_apollo_comparison}
\yuntianyihl{
    Autoware shows a higher ratio of Path Planning Optimization (PPO), due to its customizable Planning platform and larger Planning module, which accepts more contributions from the open-source community. Apollo emphasizes build, debug, and documentation-related fixes, reflecting its broader user base and the need for thorough debugging and integration with diverse hardware and software. Apollo's focus on Module Integration and Interaction (MII) patterns also highlights the complexity of maintaining compatibility across its components.
}
\end{finding}
\vspace{-1ex}

\yuntianyihl{

\subsection{\textbf{RQ2: \BFA Frequency}}

This research question explores the prevalence of domain-specific and domain-independent \bfas in ADS, aiming to categorize them into frequently occurring strategies and identify patterns that could guide targeted repair methods. By quantifying and analyzing these actions, we aim to discern which \bfas are most effective and frequently utilized across ADSes.

\begin{table}[ht]
\centering
\caption{\yuntianyihl{Domain-Independent \BFAs with Occurrences}}
\vspace{-0.5ex}
\begin{adjustbox}{width=\linewidth}
\begin{tabular}{|c|c|l|}
\hline
\textbf{Bug-Fix Action} & \textbf{\#N} & \textbf{Description} \\
\hline
\textbf{\makecell{Adjust Return Values}} & 187 & Add, delete, or change the return values, which is crucial for fixing incorrect output.  \\
\hline
\textbf{\makecell{Update System Config}}  & 177 & Add, remove, or change the system configuration or settings to ensure correct behavior. \\
\hline
\textbf{Fix API Misuse} & 166 & Correction of misused API functions to improve system interoperability. \\
\hline
\textbf{Add Condition Logic} & 147 & Add conditional statements to handle special cases or edge conditions in the code logic. \\
\hline
\textbf{Change Condition Logic} & 123 & Modify existing conditional logic to correct bugs or optimize decision-making pathways. \\
\hline
\textbf{\makecell{Update Comments}} & 109 & Update code comments or documentation to clarify or correct misunderstandings. \\
\hline
\textbf{\makecell{Modify Debug Handler}} & 96 & Enhance the error handling or debugging mechanisms to improve diagnostic ability. \\
\hline
\textbf{\makecell{Update Data Values}} & 84 & Adjustments to how data values are handled or calculated within the code. \\
\hline
\textbf{Modify Syntax Error} & 66 & Corrections of syntax errors or typos that may cause compilation or runtime errors. \\
\hline
\textbf{Move Statements} & 62 & Relocation of code statements for optimization or to resolve execution order issues. \\
\hline
\textbf{Modify GUI Property} & 58 & Changes or additions to graphical user interface elements or properties. \\
\hline
\textbf{Change Library Import} & 49 & Add, delete, or change the libraries importing to resolve dependency issues. \\
\hline
\textbf{Optimize Script Tools} & 46 & Improve the efficiency or functionality of scripts or tools used within the system. \\
\hline
\textbf{Update Dependency} & 42 & Add or remove project dependencies to resolve conflicts or ensure proper library linking. \\
\hline
\textbf{Modify Parameter Path} & 39 & Changes to parameter paths within functions or systems to resolve accessibility issues. \\
\hline
\textbf{\makecell{Add/Delete/Modify \\Parameter Declaration}} & 39 & Adjust variable or parameter declarations to fix scoping or initialization problems. \\
\hline
\textbf{Change Data Processing} & 31 & Modifications to how data is processed, often to fix errors or improve efficiency. \\
\hline
\textbf{Fix Memory Issue} & 30 & Resolution of memory management issues, such as leaks or improper allocation. \\
\hline
\textbf{Modify Command} & 29 & Update commands or options to resolve functional issues or improve user interaction. \\
\hline
\textbf{Change Data Type} & 28 & Changes in data types to resolve type mismatches or improve data integrity. \\
\hline
\end{tabular}
\end{adjustbox}
\label{table:di_bfa}
\end{table}

\autoref{table:di_bfa} presents the data of frequent domain-independent \bfas that occurred more than 20 times. The most common actions involve adjustments to return values (187 occurrences) and system configurations or settings (177 occurrences), which emphasize the need for fine-tuning output and system behavior. This reflects the dynamic nature of ADS, where precise control over return values and system configurations is essential for performance and safety.
API misuse fixes (166 occurrences) are also frequent, suggesting that proper API usage is critical in ADS development. Errors in API functions can lead to cascading issues in the system, necessitating frequent corrections. Additionally, logical condition changes and additions (270 occurrences in total) indicate that much of the bug-fixing effort in ADS involves improving decision-making logic, highlighting the complexity of handling edge cases and optimizing pathways.

The frequency of certain \bfas, such as return value adjustments and API misuse corrections, suggests that developing specialized automated repair tools for these categories could substantially reduce manual debugging efforts. Additionally, the prominence of logical condition modifications implies that enhancing logic-checking mechanisms could improve overall system reliability and performance.
By identifying recurring actions, this study provides actionable insights for prioritizing repair strategies that target the most frequent and critical bug types in ADS.

\vspace{-1ex}
\begin{finding}
\label{finding:rq2_di_bfa}
\yuntianyihl{
    The most frequent \bfas in ADS are related to adjustments of return values and configurations, indicating the critical need for precise control over output and system behavior. Frequent API misuse fixes also highlight the importance of proper API usage, as errors can propagate and cause system-wide issues. Logical condition changes and additions point to the complexity of decision-making logic in ADS, emphasizing the need to handle edge cases.
}
\end{finding}
\vspace{-1ex}
}

\begin{figure}[ht]
    \centering
    \noindent\makebox[\linewidth]{\includegraphics[width=0.8\linewidth]{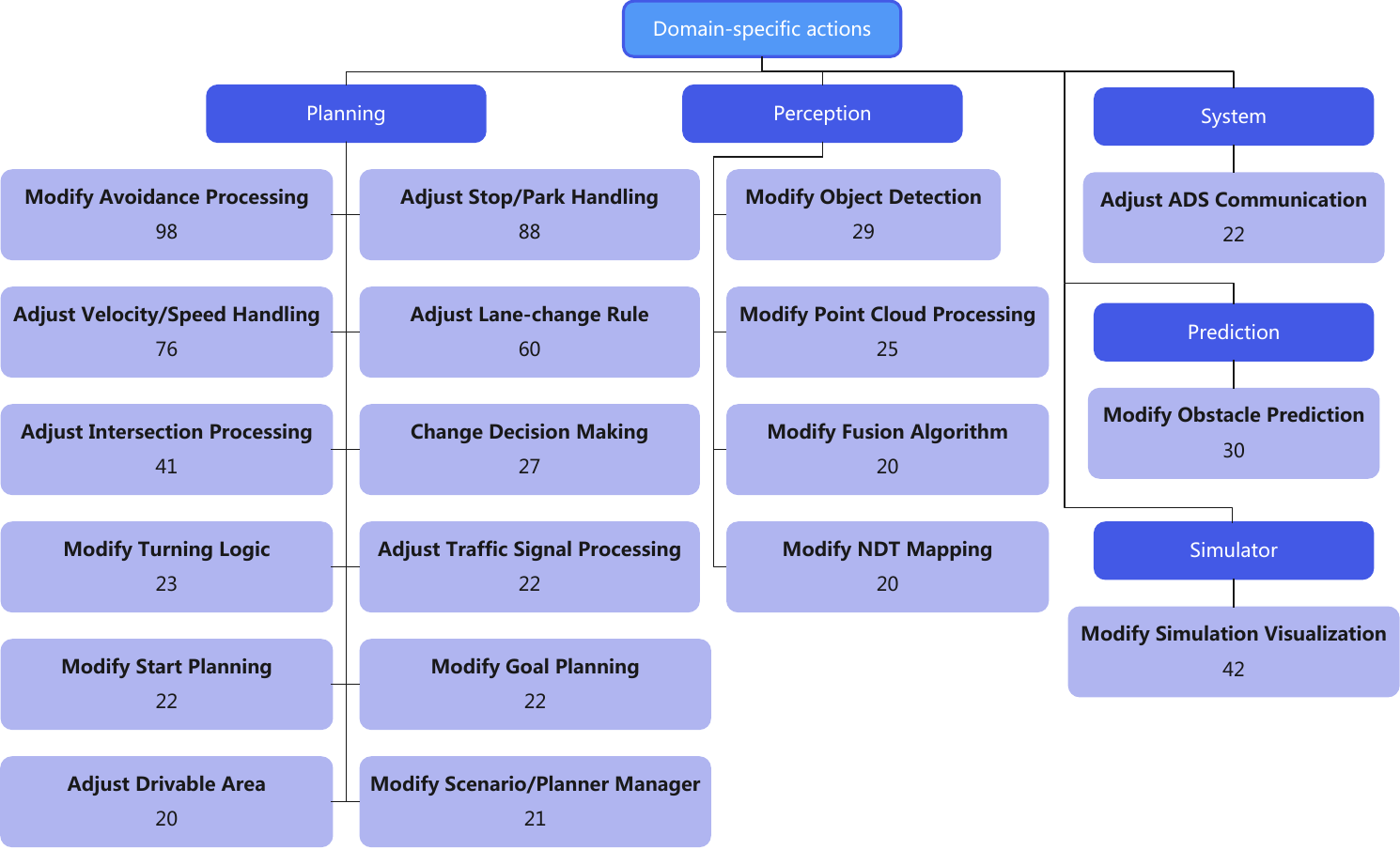}}
    \caption{\yuntianyihl{Domain-Specific \BFAs with Occurrences}}
    \label{fig:ds_bfa}
\end{figure}

As shown in \autoref{fig:ds_bfa}, we also examine the frequency of domain-specific \bfas in ADSes, focusing on the most frequent actions that occur more than 20 times and highlighting their importance within the ADS context. 
The domain-specific data clearly illustrates that Planning is the most critical ADS module, with a diverse range of sub-modules contributing to the system's overall functionality and 12 frequent \bfas, such as \textit{Modify Avoidance Processing} (98 occurrences) and \textit{Adjust Stop/Park Handling} (88 occurrences), point to its centrality in ADS operations. Planning sub-modules are integral to both path planning and lane-change processing, such as obstacle avoidance and velocity handling, reflecting their crucial role in vehicle decision-making. In addition to Planning, the Perception module stands out as the second most important and complex module of the ADS, with 4 frequent \bfas. The Perception module plays a critical role in integrating data from object detection sensors and software fusion algorithms, making it essential for accurately interpreting the environment. The complexity of this module lies in its dual functionality—relying on both hardware (sensors) and software (fusion algorithms)—to ensure real-time object detection, mapping, and environmental understanding within the ADS.

Object and obstacle processing emerges as one of the most important ADS functionalities, with \bfas like 98 occurrences of \textit{Modify Avoidance Processing}, 30 of \textit{Modify Obstacle Prediction}, and 29 of \textit{Modify Object Detection} being highly frequent. This demonstrates that ensuring system's ability to detect, predict, and avoid obstacles is a top priority. These actions span multiple ADS modules, including Perception (object detection), Planning (avoidance), and Prediction (obstacle prediction), highlighting the complexity and interdependence of object handling across the ADS.

\vspace{-1ex}
\begin{finding}
\label{finding:rq2_obs}
\yuntianyihl{
    Object and obstacle processing is a major focus, 
    emphasizing the importance of detecting, predicting, and avoiding obstacles. The frequent occurrence of object-processing actions across multiple ADS modules reflects the interdependence and complexity of handling obstacles, highlighting the priority of maintaining robust object-handling mechanism.
}
\end{finding}
\vspace{-1ex}

Autonomous vehicle maneuvers play an important role in ADS path planning. Frequent \bfas like \textit{Adjust Stop/Park Handling} (88 occurrences), \textit{Adjust Velocity/Speed Handling} (76 occurrences), and \textit{Modify Turning Logic} (23 occurrences) stress the importance of refining these maneuvers to avoid accidents and ensure smooth navigation. The complexity of handling speed and velocity, in particular, indicates a need for ongoing refinement in dynamic vehicle control.

Lane-related \bfas also feature prominently, with \textit{Adjust Lane-change Rule} (60 occurrences), \textit{Modify Goal Planning} (22 occurrences), \textit{Modify Start Planning} (22 occurrences), and \textit{Adjust Drivable Area} (20 occurrences) indicating the system's reliance on precise lane generation and start and destination planning for safe and efficient travel. Lane changes, drivable area processing, and goal planning are all essential for adaptive route selection, further emphasizing the importance of these components within the broader ADS planning framework.

The \textit{Modify Simulation Visualization} (42 occurrences) demonstrates the critical need for simulator reliability in ADS development. Simulations are used to test and verify system behaviors before deployment in the real world. A stable and accurate simulation environment is vital for scenario-based testing, such as obstacle detection and collision avoidance, to ensure that the ADS performs as expected in diverse conditions. High-frequency modifications in this area signal the need to continuously refine simulation environments for more reliable and user-friendly ADS simulators.

\vspace{-1ex}
\begin{finding}
\label{finding:rq2_combined}
    Frequent maneuver-related bug-fix actions (187 occurrences) highlight the need for precise control to ensure safe navigation and prevent accidents. Lane-related actions (102 occurrences) underscore the importance of accurate lane and route planning for adaptive and efficient travel. The simulation-related action (42 occurrences) emphasizes the critical role of simulation in verifying ADS behaviors under various conditions. Frequent adjustments to the simulator highlight the ongoing need to enhance its stability and accuracy for reliable testing.
\end{finding}
\vspace{-1ex}

\subsection{\textbf{RQ3: \BFPs Across Root Causes}}

\begin{figure}[ht!]
\begin{minipage}{0.66\textwidth}
\captionof{table}{Classification of \yuntianyihl{Affected} Algorithm \yuntianyihl{Groups} with \yuntianyihl{Frequency}}
\begin{adjustbox}{width=\linewidth}
\begin{tabular}{c|l|c|l}
\toprule
\textbf{Algorithm Group} & \textbf{Subgroup} & \textbf{Total (\%)} & \textbf{Example} \\
\midrule
\multirow{6}{*}{\textbf{\makecell{Planning \&\\ Navigation}}} & Route planning and goal setting & 41 (5.5\%) & Start Planner \\
 & Lane-related maneuvers & 54 (7.2\%) & Lane Change \\
 & Evasive actions & 82 (11.0\%) & Avoidance \\
 & Parking and stopping maneuvers & 25 (3.4\%) & Pull Over \\
 & Intersection handling & 40 (5.4\%) & Scene Intersection\\
 & Route management & 44 (5.9\%) & Mission Planner \\
\midrule
\multirow{2}{*}{\textbf{\makecell{Control \&\\ Actuation}}} & Motion control & 18 (2.4\%) & MPC Controller \\
 & Command decision systems & 18 (2.4\%) & Turn Signal Decider \\
\midrule
\multirow{3}{*}{\textbf{\makecell{Perception \&\\ Sensing}}} & Sensor data processing & 3 (0.4\%) & Pointcloud Preprocessor \\
 & Sensor data fusion & 6 (0.8\%) & Perception Fusion \\
 & Object and environment detection & 30 (4.0\%) & CV Tracker \\
\midrule
\multirow{2}{*}{\textbf{Prediction}} & Entity Behavior Prediction & 12 (1.6\%) & Map Based Prediction \\
 & Traffic and Environmental Prediction & 3 (0.4\%) & Traffic Light Predictor \\
\midrule
\multirow{2}{*}{\textbf{\makecell{Mapping \&\\ Localization}}} & Map handling and management & 34 (4.6\%) & PNC Map \\
 & Localization and position estimation & 8 (1.1\%) & EKF Localizer \\
\midrule
\multirow{2}{*}{\textbf{\makecell{System Management \\\& Infrastructure}}} & System monitoring and management & 24 (3.2\%) & System Error Monitor \\
 & Data management and configuration & 9 (1.2\%) & CSV Loader\\
\midrule
\multirow{2}{*}{\textbf{\makecell{Simulation, Testing,\\ \& Evaluation}}} & Simulation and testing environments & 7 (0.9\%) & Planning Simulator \\
 & Performance evaluation and analysis & 7 (0.9\%) & Trajectory Evaluator \\
\midrule
\multirow{2}{*}{\textbf{\makecell{User Interface \&\\ Communication}}} & HMI and display systems & 4 (0.5\%) & RTC Manager Panel \\
 & Communication and data exchange & 2 (0.3\%) & Cyber Channel \\
\midrule
\multirow{2}{*}{\textbf{\makecell{Emergency \\Handling \& Safety}}} & Emergency response & 10 (1.3\%) & Emergency Handler \\
 & Risk checking and safety planning & 3 (0.4\%) & Path Safety Checker \\
\bottomrule
\end{tabular}
\end{adjustbox}
\label{tab:algorithm_classification}
\end{minipage}
\hfill
\begin{minipage}{0.33\textwidth}
    \vspace{1ex}
    \centering
    \includegraphics[width=\linewidth]{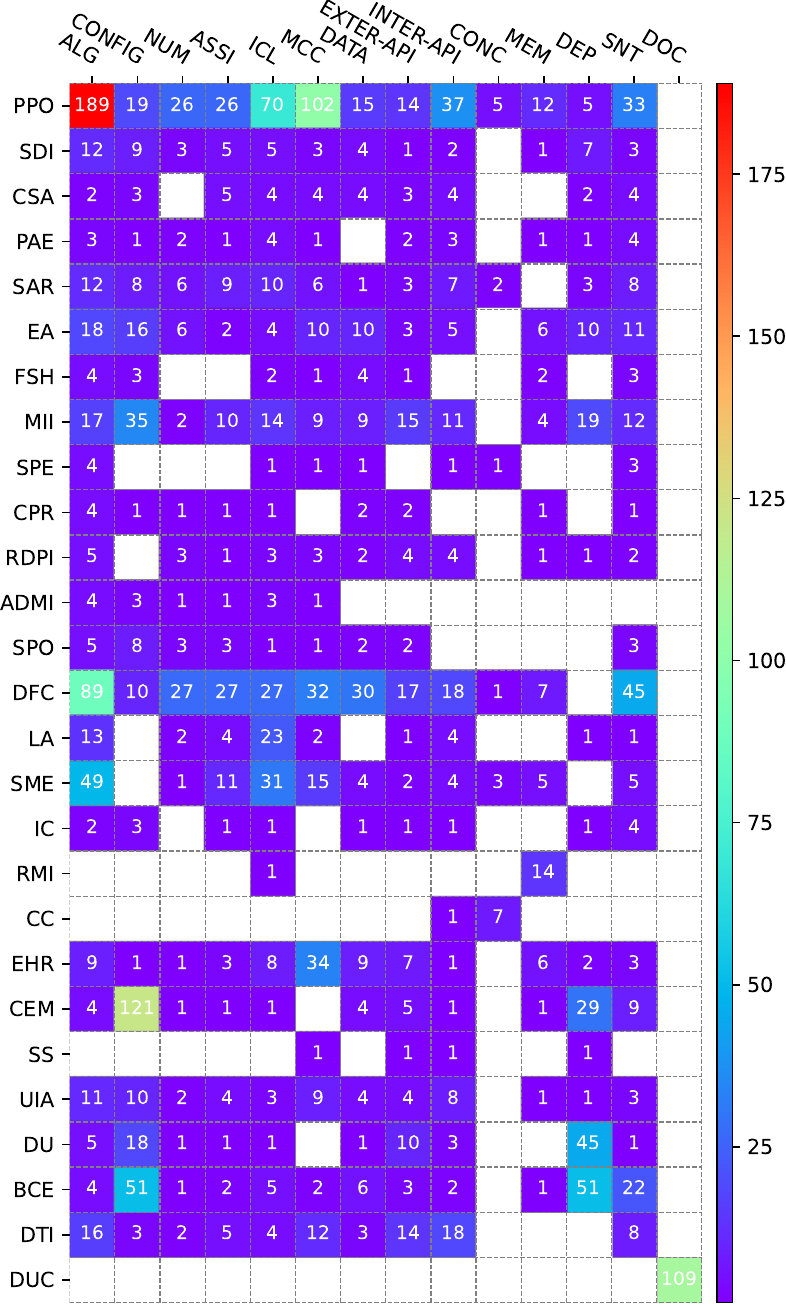}
    \vspace{-5ex}
    \captionof{figure}{Relation between root causes and Semantic \yuntianyihl{BFPs} for both ADSes}
    \label{fig:rootcause_semantic_combined}
\end{minipage}
\end{figure}

For both ADSes, we identified a total of 1,331 instances of these root causes.
~\autoref{fig:rootcause_semantic_combined} 
correlate semantic BFPs with root causes. 
\yuntianyihl{The Algorithm (Alg) root cause has a significant occurrence of 887 bug fixes and spans across 23 types of semantic \bfps, which}
underlines the complexity and multifaceted nature of these bugs.
This observation, aligned with the detailed classification of the affected algorithm or functions for pull requests in \hyperref[tab:algorithm_classification]{Table 8}, where \textit{Evasive Actions}, particularly related to obstacle processing, emerges as the most corrected subgroup, underscoring the criticality of refining obstacle avoiding algorithms in ADS.
The dataset we collected presents a compelling case for developing learning-based automated program repair (APR) approaches, particularly as the top five affected algorithms---\textit{Avoidance}, \textit{Lane Change}, \textit{Scene Intersection}, \textit{Start Planner}, and \textit{Goal Planner}---point to specific areas where such APR strategies could be most beneficial.

\vspace{-1ex}
\begin{finding}
\label{finding:rq3_alg}
        Algorithm-related bug fixes, with a total of 877 occurrences, are characterized by various semantic patterns that require multifaceted code modifications to address the underlying incorrect algorithm logic. This complexity, notably present in the \textit{Evasive Actions} subgroup, suggests an opportunity for advancing APR techniques for obstacle processing algorithms.
\end{finding}
\vspace{-1ex}

For the Config root cause, 
121 occurrences of Configuration and Environment Management (CEM), 51 of Build and Compilation Enhancement (BCE), and 35 of Module Integration and Interaction (MII) dominate semantic patterns. This trend suggests that configuration-related bugs not only require changes in system configuration files but also demand adjustments in build files and hardware settings. This highlights the broader scope of configuration bugs, which extend beyond code to include systemic and environmental factors.
According to recent surveys in automated program repair~\cite{apr_survey,ZhangFMSC24}, although APR has made strides in addressing code-level issues, the aspect of configuration repair remains relatively unexplored, necessitating focused efforts to fill this gap.

\vspace{-1.5ex}

\vspace{-1ex}
\begin{finding}
\label{finding:rq3_config}
    Most configuration-related bugs are fixed by 
    Configuration and Environment Management (CEM), Build and Compilation Enhancement (BCE), and Module Integration and Interaction (MII) patterns (207 occurrences in total),
    indicating the need for comprehensive fixes that span both source code and configurations. 
    This situation emphasizes the critical gap in automated testing and program repair research concerning ADS configuration.
\end{finding}
\vspace{-1ex}

\subsection{\textbf{RQ4: \BFPs Across Symptoms}}

\begin{figure}[ht!]
    \centering
    \begin{minipage}{0.525\linewidth}
        \centering
        \includegraphics[width=\linewidth]{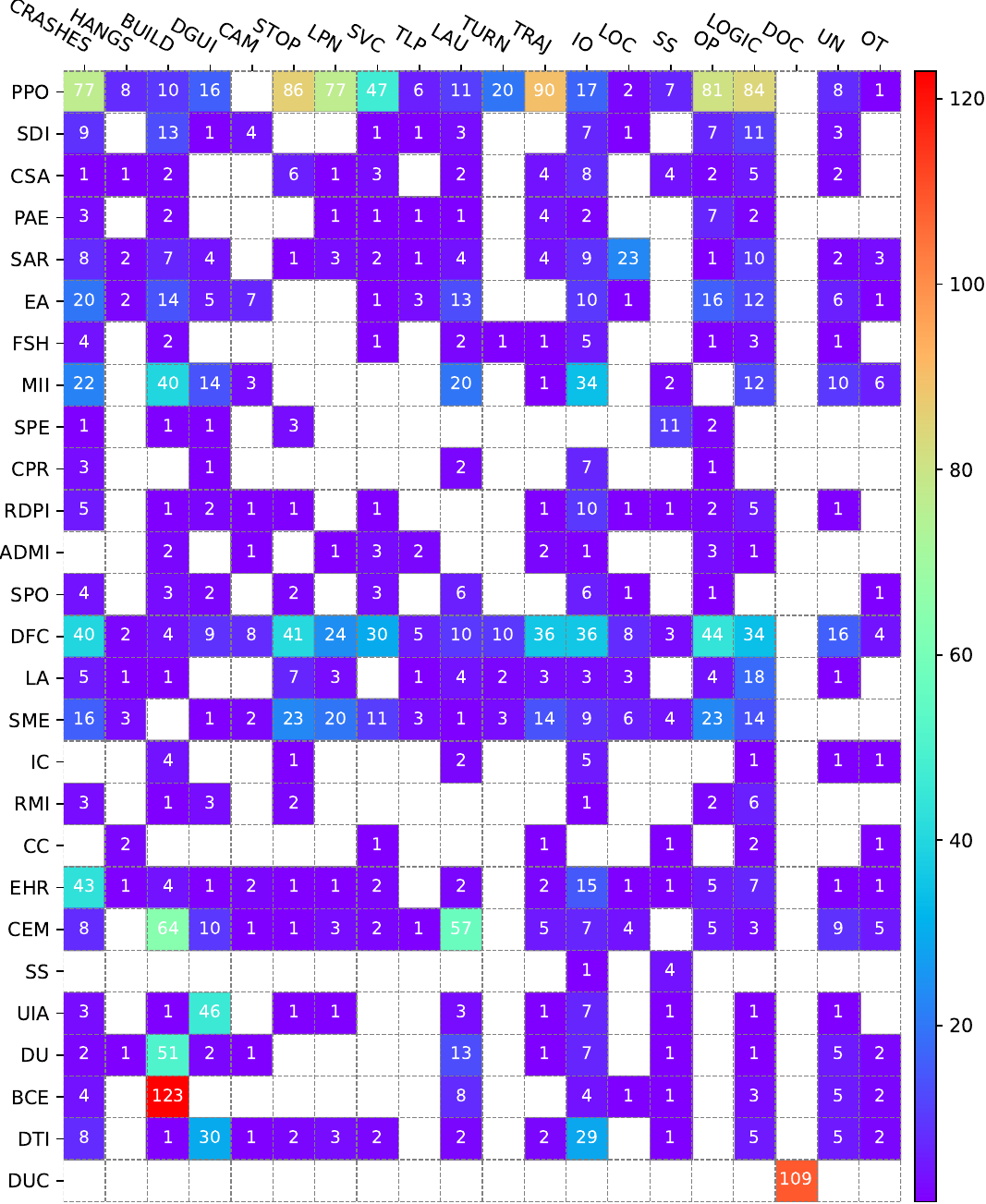}
        \vspace{-5ex}
        \caption{Relation between Symptoms and Semantic \BFPs for Apollo and Autoware}
        \label{fig:symptom_semantic_combined}
    \end{minipage}
    \hfill
    \begin{minipage}{0.465\linewidth}
        \centering
        \includegraphics[width=\linewidth]{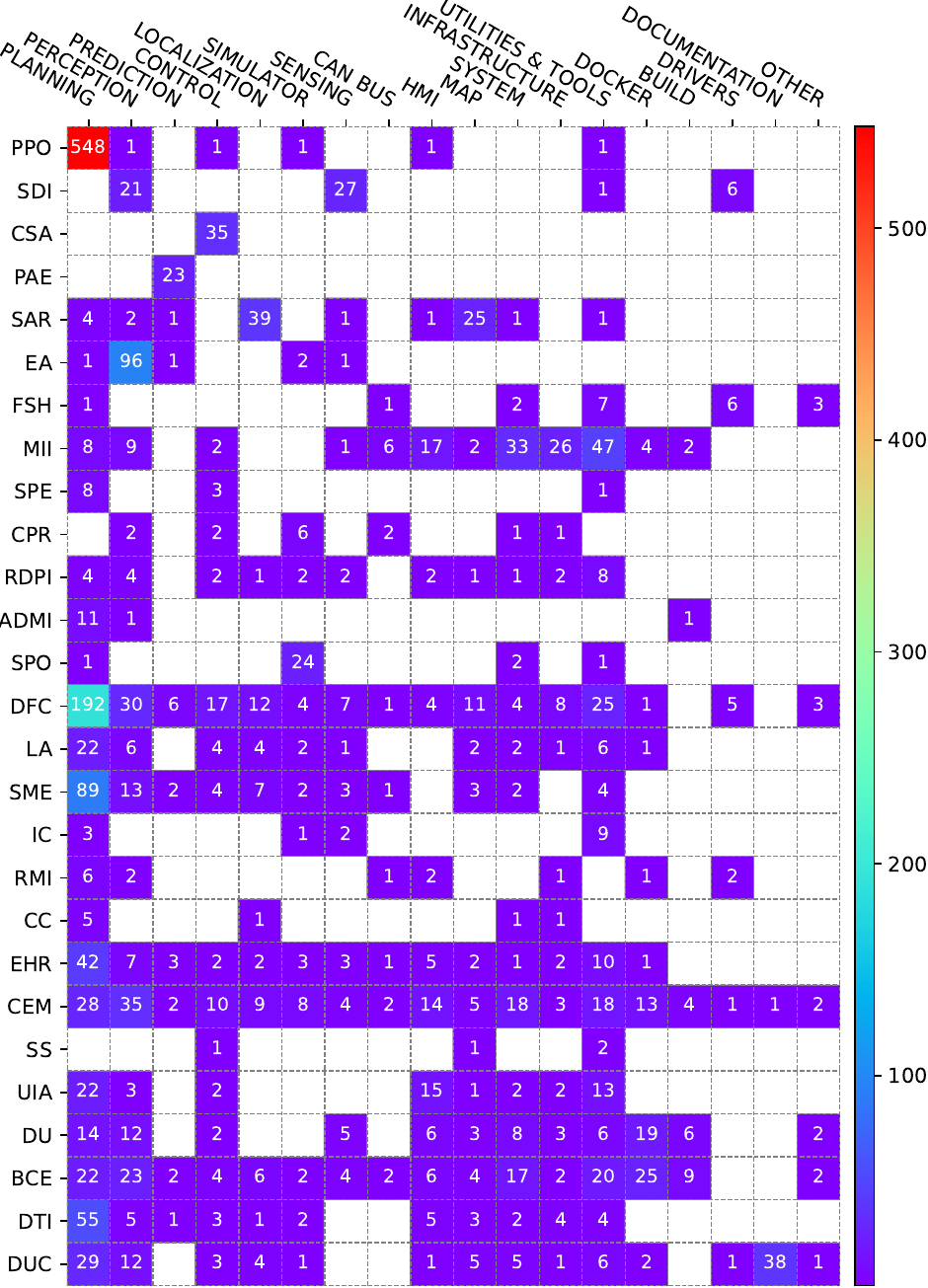}
        \vspace{-5ex}
        \caption{Relation between Modules and Semantic \BFPs for Apollo and Autoware}
        \label{fig:module_semantic_combined}
    \end{minipage}
\end{figure}

As shown in \autoref{fig:symptom_semantic_combined}, this research question illustrates the relationship between bug symptoms observed in Apollo and Autoware and semantic BFPs, with 2473 symptom instances in total.

The occurrence of domain-specific \bfp Path Planning Optimization (PPO) and domain-independent pattern Data Flow Correction (DFC) across various symptoms, such as Crashes, Stop and parking (Stop), Lane Positioning and Navigating (LPN), Speed and Velocity Control (SVC), Turning (Turn), Trajectory (Traj), Obstacle Processing (OP), and Logic, signifies the critical nature of these patterns in addressing diverse and complex issues in different driving scenarios. 

\vspace{-1ex}
\begin{finding}
\label{finding:rq4_dfc}
    The frequent occurrence of Data Flow Correction (DFC) (364 occurrences) in diverse symptoms underscores the importance of data-flow handling for reliable autonomous vehicle operations, which suggests the need for robust data-flow testing and correction for ADSes.
\end{finding}
\vspace{-1ex}

The 30 instances of Debugging Tools Improvement (DTI) bug fixes, primarily in Display and GUI (DGUI) symptoms, demonstrate that ADS developers value debugging methods that are specialized to their domain, but further value effective visualizations for debugging concerns. 
Existing debugging techniques~\cite{ArcainiCILZAHV21,MINNERUP201644,apollo_debug,autoware_debug}, such as log analysis, breakpoint debugging, and unit testing, provide a broad foundation for software diagnosis. However, the absence of dedicated research on convenient debugging tools for ADS reveals a critical gap, where the intricate and real-time nature of ADS presents unique challenges such as complex sensor data integration, safety-critical decision-making, and dynamic environmental interaction. Improved visualization tools for debugging may consider offering a more intuitive understanding of ADS behavior and error states.

\vspace{-1ex}
\begin{finding}
\label{finding:rq4_dti}
    The prevalence of 30 Debugging Tools Improvement (DTI) bug fixes in DGUI symptoms highlights the critical need for ongoing improvements in debugging tools and user interfaces within autonomous vehicle systems. 
\end{finding}
\vspace{-1ex}

State Management Enhancement (SME) \bfps are most commonly observed in Stop and Parking (Stop) (23 occurrences), Obstacle Processing (OP) (23 occurrences), and Lane Positioning and Navigating (LPN) (20 occurrences) symptoms. 
Effective state management is crucial for ensuring seamless transitions between different operational modes, such as transitioning from driving to parking or navigating through obstacles. This necessity points to the intricate interplay between state control mechanisms and the vehicle's capacity to interpret and react to dynamic driving environments, thereby stressing the importance of sophisticated state handling strategies to mitigate potential risks and enhance driving safety.
These scenarios, which necessitate precise state transitions such as moving from forward to reverse for parking or adjusting to dynamic obstacles, highlight the critical role of advanced state management.

\vspace{-1ex}
\begin{finding}
    \label{finding:rq4_sme}
    The concentration of State Management Enhancement (SME) patterns in scenarios like Stop and Parking, Obstacle Processing, and Lane Positioning and Navigating underscores the importance of robust state management for AV adaptability to diverse driving conditions. 
\end{finding}
\vspace{-1ex}

The significant count of 178 STOP-related \bfps in AVs underscores the complex and persistent issues in stopping and parking functionalities, revealing systemic vulnerabilities and the need for refined parking algorithms and community concerns. These patterns suggest that current methodologies may fall short in addressing the intricate interactions among subsystems essential for efficient stop and parking operations. Few existing papers focus on enhancing the parking and stopping of the ADS~\cite{IbischSANTSSK13,BanzhafNKZ17}, this gap in research and development highlights the urgency of advancing testing frameworks and enhancing safety standards to address the operational complexities and potential hazards inherent in AV stop and parking mechanisms.

\vspace{-1ex}
\begin{finding}
\label{finding:rq4_stop}
    The presence of 178 STOP-related \bfps and few AV studies focusing on such functionalities reveals a major gap of \yuntianyihl{stop and parking optimization} in ADS research.
\end{finding}
\vspace{-1ex}

\subsection{\textbf{RQ5: \BFPs Across ADS Modules}}

This RQ investigates the degree of repair efforts associated with ADS modules. 
\autoref{fig:module_semantic_combined} presents the occurrences of semantic \bfps within distinct modules of Apollo and Autoware.

\begin{wrapfigure}[15]{r}{0.52\textwidth}
    \centering
    \vspace{-2ex}
    \noindent\makebox[\linewidth]{\includegraphics[width=\linewidth]{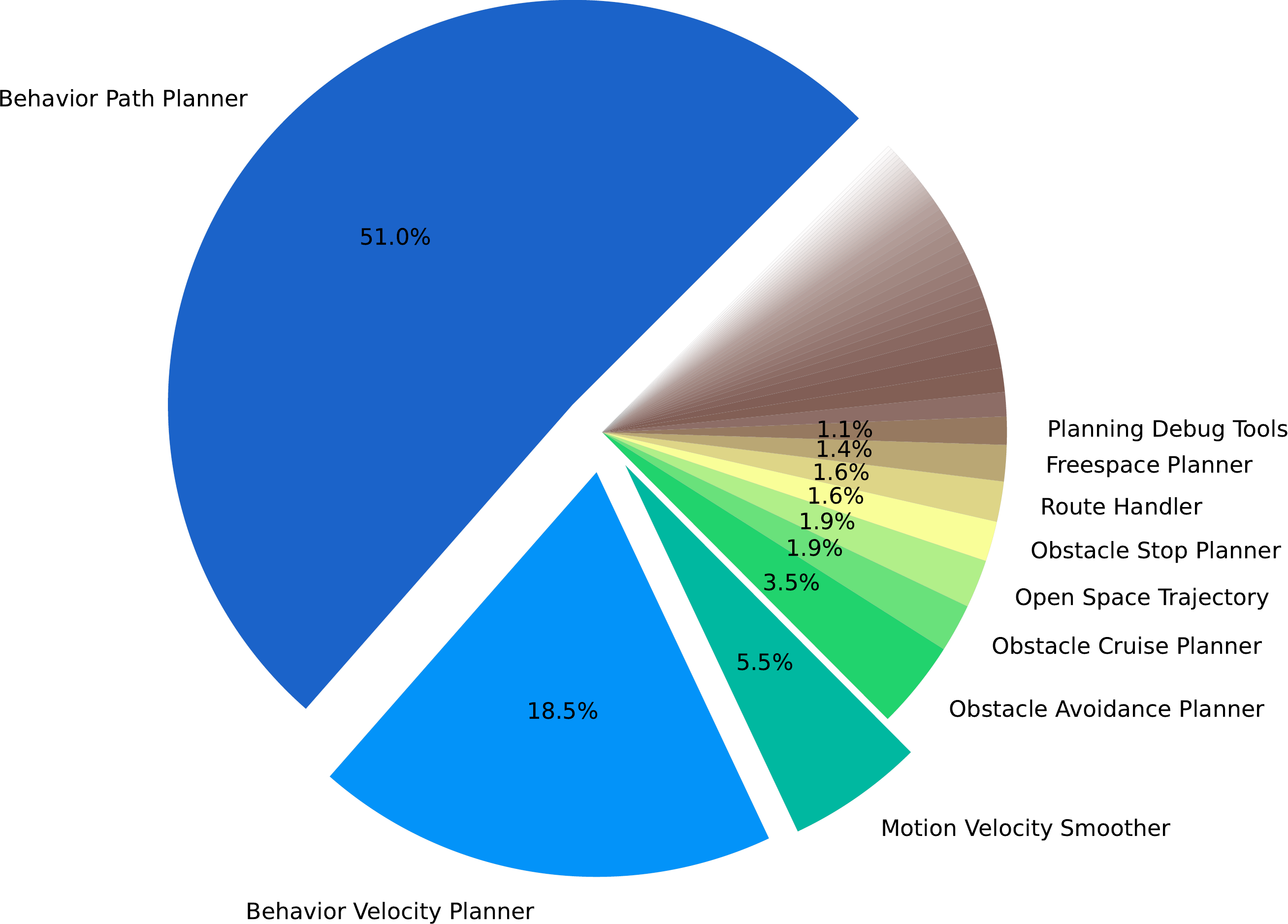}}
    \vspace{-4ex}
    \caption{Distribution of Sub-modules in Planning Module (Labels less than 1.0\% are hidden)}
    \label{fig:pie_subcomponent_combined}
\end{wrapfigure}

In the ADS domain, the Planning module is distinguished by its coverage of 22 types of semantic BFPs, with PPO manifesting 548 times, making it one of the most frequently occurring patterns in ADS modules.
The Planning module is also the most complicated module, with a total of 50 sub-modules for Apollo and 51 for Autoware. With its diverse sub-modules like trajectory generation and velocity planning, the Planning's complexity and centrality in ADS underscore the necessity for improved design and modularization for enhanced maintenance and functionality.

The Planning's intricacy is further illuminated by the frequent occurrence of PPO and its subdivision into numerous sub-modules, as highlighted in \autoref{fig:pie_subcomponent_combined}, 
\yuntianyihl{such as the \textit{Behavior Path Planner} (51.0\%), whose dominance signifies its critical role in determining the optimal path. Its frequent interaction with other subsystems makes it a focal point for potential errors. The \textit{Behavior Velocity Planner} at 18.5\% further emphasizes the complexity of real-time decision-making in velocity adjustments, affecting safety and performance. \textit{The Motion Velocity Smoother} at 5.5\% plays a crucial role in ensuring the vehicle's movements remain smooth and predictable, reducing jerky motions that could compromise passenger comfort and safety.}
This intricate structure \yuntianyihl{of interdependent sub-modules} and the transition of Apollo to fragment modules into smaller packages in \textit{v9.0}~\cite{apollo_architecture, apollo_v9} reflects the growing need for refined architectural and modularization approaches in ADS, particularly to enhance the Planning's efficiency and maintainability \yuntianyihl{in handling real-world challenges like obstacle avoidance and smooth velocity transitions. The Planning's complexity, as evidenced by the high frequency of semantic bugs in its core sub-modules, indicates that further research into modularization could lead to significant performance improvements and error reduction.}

\vspace{-1ex}
\begin{finding}
\label{finding:rq5_planning}
    The prominent occurrence of 548 instances of semantic \bfps and about 50 sub-modules in the Planning module underscores its centrality and complexity in both ADSes. 
    \yuntianyihl{Key sub-modules about path, velocity, and avoidance planning contribute significantly to this complexity due to their high functional load and frequent interaction with other sub-modules. The varied roles of these sub-modules highlight the need for a more refined software architecture and improved granular modularization to facilitate easier maintenance and system upgrades.}
\end{finding}
\vspace{-1ex}

Domain-specific semantic BFPs are predominantly associated with particular modules, whereas domain-independent BFPs are more evenly spread across various modules. For instance, SLAM Algorithm Refinement (SAR) is specific to the Localization and Environmental Adaptability (EA) to the Perception. 
This distribution implies domain-specific issues are closely tied to the functionalities of their respective modules, reflecting adherence to the principle of single responsibility~\cite{martin2000design}, where each module is closely tied to distinct functionalities, necessitating precise, module-specific solutions. 
Conversely, domain-independent patterns, embodying cross-cutting concerns~\cite{crosscutting_concerns,EaddyZSGMNA08}, pervade various modules, suggesting a need for overarching strategies that address these broader issues.

\vspace{-1ex}
\begin{finding}
\label{finding:rq5_specific_independent}
    While domain-specific bug-fix patterns tend to coalesce to their most semantically similar ADS module, domain-independent ADS bug-fix patterns cross-cut modules. 
\end{finding}
\vspace{-1ex}

Configuration and Environment Management (CEM) is present across all modules, with a notable prevalence in the Perception module (35 occurrences). This indicates that configuration and environmental factors are universally important across modules, with the Perception module being particularly sensitive to these aspects. The pervasive nature of CEM patterns suggests that consistent and accurate configuration and environment management are critical for the seamless operation of autonomous driving systems, especially in processing and interpreting sensory data.

Data Flow Correction (DFC), with a total of 330 occurrences in all modules, stands out as the most common domain-independent \bfp, indicating that issues related to data handling and processing are pervasive across various modules. 
This prevalence suggests that ensuring the accuracy and efficiency of data flow is a critical concern in ADSes, reflecting the need for robust mechanisms to manage and correct data flow across the system's diverse components. 

\vspace{-1ex}
\begin{finding}
\label{finding:rq5_dfc}
    The prominence of Data Flow Correction (DFC) as a domain-independent BFP highlights the critical need for accurate and efficient data flow management across the diverse components of ADSes. This prevalence indicates a pressing demand for advanced APR heuristics and debugging tools specifically designed to address the complex data-flow challenges in ADS.
\end{finding}
\vspace{-1ex}

    \section{Discussion}\label{sec:discussion}

The prevalence of Data Flow Correction (DFC) underscores the pivotal role of accurate data-flow 
handling in the 
ADS (Findings \ref{finding:rq1_dfc}, \ref{finding:rq4_dfc},
and \ref{finding:rq5_dfc}). 
DFC's significance across various operational contexts reveals that data-flow issues are a major factor in system bugs, necessitating advanced testing and repair strategies. 
The need for improved tracking and visualization of 
data-flow
is further backed by our study's results, as ADS engineers appear to already struggle from these issues (\autoref{finding:rq4_dti}).

In terms of software testing, our study reveals the need for a greater focus on ADS module integration testing (\autoref{finding:rq1_ppo_mii}) and for more general software testing focused on understudied scenario types, e.g., AV stopping and parking (Findings \ref{finding:rq4_sme} and \ref{finding:rq4_stop}). In fact, the many sub-modules of Planning indicate that there are a wide variety of opportunities to potentially handle many sub-functionalities of Planning and test them in tandem (e.g., obstacle processing and avoidance (Findings \ref{finding:rq2_obs} and \ref{finding:rq3_alg}) and parking (Findings \ref{finding:rq2_combined} and \ref{finding:rq4_sme})) and with other modules (e.g., Planning and Control along the interactions between software, hardware, and the physical environment (\autoref{finding:rq1_ppo_mii})).

The manner in which ADS domain-specific BFPs tend to coalesce in semantically similar ADS modules, while domain-independent BFPs cross-cut those modules (\autoref{finding:rq5_specific_independent}), along with certain ADS modules (e.g., Planning) being architected into many sub-modules (\autoref{finding:rq5_planning}), suggests that ADS engineers highly value properly-encapsulated and well-architected ADS modules. We are not aware of a previous study that correlated bug fixing and software architecture in such a manner. Future research may consider whether machine learning-based modules (e.g., Prediction and Perception) may benefit in terms of bug fixing from explicit modularization and encapsulation.

\yuntianyihl{The dataset in this study could serve as a benchmark to facilitate future automated program repair research in the ADS domain. 
The high occurrences of certain \bfas, such as \textit{Adjust Return Values} and \textit{Fix API Misuse} (\autoref{finding:rq2_di_bfa}), suggest that developing specialized automated repair tools for these categories could substantially reduce manual debugging efforts. Additionally, the prominence of logical condition modifications implies that enhancing logic-checking mechanisms could improve overall system reliability and performance.
By identifying recurring actions, this study provides actionable insights for prioritizing repair strategies that target the most frequent and critical bug types in ADS.
}

    \section{Threats to Validity}\label{sec:threats}

\noindent \textbf{Internal Threats.} 
The main internal threat to the validity of this study is the potential subjective bias or errors in bug classification. To mitigate this, we began our labeling process using established classification schemes from the literature~\cite{ThungWLJ12, SeamanSREFGG08} and employed an open-coding scheme~\cite{blair2015reflexive} to expand these initial frameworks. We also concentrated on real bug fixes by selecting only closed and merged pull requests. Each bug was independently examined and labeled by two ADS developers, both of whom have contributed to Apollo and Autoware, with any discrepancies resolved through discussion until consensus was achieved. In addition, discussions with Apollo and Autoware developers were held to refine our classification, further reducing subjective biases and errors.

Additionally, our approach to identifying bugs 
relies heavily on keywords 
in the context of pull requests. This method might result in false positives since developers sometimes mistakenly mark feature additions as bug fixes.
We mitigated this by filtering out false positives during labeling.

\noindent \textbf{External Threats.} 
The primary external threat concerns the reliability of our dataset. To address this, our data collection includes all pull requests, commits, and issues from the inception of the subject ADSes until the major version release at the end of 2023. 
This comprehensive dataset aligns with similar methods~\cite{GarciaF0AXC20,IslamNPR19,VasilescuYWDF15,ZhangCCXZ18,FrancoGR17} used in other bug studies to identify bug-fix pull requests. 
Another threat is the open-source nature of the ADSes under study. Our study involves two widely-used representatives---Apollo and Autoware, developed independently and featuring extensive code bases, which may not reflect the bug-fix patterns prevalent in closed-source environments due to differing development processes. 
\yuntianyihl{As for the generalizability, despite the study encompassing only two systems, they have significant industry and research usage, including the US government, Google, Ford, and various car manufacturers~\cite{40_plus_corporations,baidu_volvo_ford,baidu_apolong,carma_github,18_av_companies,carma_overview,waymo_8_miles,lyft_ces}, and have been developed, iterated, and studied for years, accumulating more research studies, issues, and pull requests than any other small-scale open-source ADS projects. 
\yuntianc{The two projects we selected represent the forefront of open-source ADS development.}
Furthermore, most ADSes, including Apollo and Autoware, share the similar reference architecture and are developed from the ROS architecture~\cite{ros_icra}.}

Besides, the number of labeled bugs in our study (\numbug) is comparable to other \bfp studies in different domains (e.g., 970 in deep neural networks~\cite{IslamPNR20}, 446 in deep learning stack~\cite{Huang0WCM023}, and 395 in federated learning systems~\cite{DuCC0C023}), underscoring the representativeness of our findings.

    \section{Related Work}\label{sec:related}

\noindent \textbf{ADS Bug Study.} ADSes have been a focus of recent software research, particularly concerning software bugs. Garcia et al.~\cite{GarciaF0AXC20} provided a comprehensive study of bugs in AVs, which has become increasingly relevant as these systems become more prevalent in the automotive industry. 
Building on this, Zampetti et al.~\cite{ZampettiKPP22} conducted an empirical characterization of software bugs in open-source cyber-physical systems, which include ADS. Their study contributes to the understanding of the challenges and common issues in the development and maintenance of these complex systems.

\noindent \textbf{Bug-Fix Patterns.} Syntactic \bfps primarily involve changes in the code structure and statements.
Pan et al.~\cite{PanKW09} were among the pioneers in exploring this area, providing foundational insights into the nature and taxonomy of syntactic bug fixes. This study laid the groundwork for subsequent research, including Zhong et al.~\cite{ZhongS15}, who conducted an empirical study on real bug fixes, focusing on the features of Java projects. 
Soto et al.~\cite{SotoTWGL16} looked deeper into bug fixes by examining patterns, replacements, deletions, and additions. 
Following this, Campos et al.~\cite{CamposM17,CamposM19} conducted large-scale observational studies to identify common BFPs, which were instrumental in cataloging the recurrent patterns in bug fixes across various software.
Additionally, Islam et al.~\cite{IslamZ20} introduced more types of syntactic BFPs based on the study of Pan et al.~\cite{PanKW09} and provided an in-depth analysis of BFPs, particularly focusing on the structural edits and nesting levels in code. 

For semantic \bfps, Islam et al.~\cite{IslamPNR20} explored the bug fixes in deep neural networks. 
Huang et al.~\cite{Huang0WCM023} presents a comprehensive study on characterizing symptoms, root causes, and fix patterns of dependency bugs across the deep learning stack.
Du et al.~\cite{DuCC0C023} complements empirical bug studies in deep learning systems by studying six popular federated learning systems.

    \section{Conclusion}\label{sec:conclusion}

In the rapidly evolving landscape of autonomous driving systems, ensuring the reliability and safety of these systems is essential. Our empirical study, which focuses on \bfps in two major autonomous driving systems, Apollo and Autoware, has provided a comprehensive insight into the hierarchy of ADS bugs and their fix patterns in this domain. 
The dominant \bfps identified, encompassing issues related to path planning optimization, data flow correction, and if-related statements, underscore the multifaceted challenges developers face in maintaining and improving these systems.
Our proposed taxonomies of \numsyn syntactic and \numsem semantic \bfps 
serve as foundational resources for developers and researchers in the ADS domain by identifying \yuntianyihl{and providing a benchmark of} \numbug bug fixes. 
In future work, we aim to use the results and dataset we collected from the \bfp study for facilitating automatic program repair, enhancing debugging tools, training targeted developers, and cross-domain applications in the ADS domain.

    \section{Data Availability}
\label{sec:data_availability}

The dataset and additional analytical figures are available at \cite{bfp_artifacts}.

    \bibliographystyle{ACM-Reference-Format}
    \bibliography{main}
\end{document}